\documentclass[10pt]{article}
\usepackage{amssymb}
\usepackage{amscd}
\usepackage{latexsym}
\usepackage{boldtensors}
\usepackage{amsmath}
\usepackage{amsthm}
\usepackage{graphicx}
\newcommand{\newsec}{\setcounter{equation}{0}\section}%
\newcommand{\Zz}{{\mathbb Z}}
\newcommand{\Nn}{{\mathbb N}}
\newcommand{\Rr}{{\mathbb R}}

\def\be {\begin{equation}}
\def\ee{\end{equation}}
\def\bea{\begin{eqnarray}}
\def\eea{\end{eqnarray}}

\def\d{{\rm d}}
\def\i{{\rm  i}}

\newtheorem*{theorem}{Theorem}%[section]
\newtheorem{lemma}{Lemma}%[section]

\theoremstyle{remark}  
\begin{document}
\title{%{\flushleft{\small {\rm Published in     }\\}}\vspace{1cm}
\large\bf Condensation
of interacting bosons\footnote{Work supported by OTKA Grant No. K128989.}}
\author{Andr\'as S\"ut\H o\\Wigner Research Centre for Physics\\P. O. B. 49, H-1525 Budapest, Hungary\\
E-mail: suto.andras@wigner.hu\\}
\date{}
\maketitle
\thispagestyle{empty}
\begin{abstract}
\noindent
In this third paper of a series that started with
arXiv:2106.10032 [math-ph] and continued with
arXiv:2108.02659 [math-ph] we show that in $d\geq 3$ dimensions at low temperatures or high densities bosons interacting via pair potentials that are both positive and positive type form permutation cycles whose length diverges proportionally with the number of particles. Based on the second-cited paper, this implies Bose-Einstein condensation.
%PACS: 61.50.Ah, 02.30.Nw, 61.50.Lt
\end{abstract}
%%%%%%%%%%%%%%%%%%%%%%%%%%%%%%%%%%%%%%%%%%%%%%
\newsec{Introduction}
During the "trente glorieuses" of equilibrium statistical mechanics the mathematical foundations of the theory of phase transitions were laid down and many beautiful results were obtained, yet the maybe hardest question remained unanswered: how to prove phase transitions in continuous space.
There seems to exist no method comparable to that of correlation inequalities, contour models or reflection positivity, so efficient for lattice systems. Some emblematic problems that wait for a solution are the vapor-liquid and the liquid-solid transition for the Lennard-Jones potential, the crystallization of hard balls and the putative hexatic transition of hard disks. Here we attack another open problem, the Bose-Einstein condensation (BEC) of interacting atoms.
\begin{theorem}
Consider $N$ identical bosons on a $d\geq 3$-torus of side $L$ at inverse temperature $\beta$ that interact via a pair potential $u:\Rr^d\to\Rr$ of the following properties:
\begin{itemize}
\item[(i)]
$u\geq 0$.
\item[(ii)]
The Fourier-transform $\hat{u}$ of $u$ exists, it is integrable and nonnegative ($u$ is of the positive type).
\item[(iii)]
$u(x)=O\left(|x|^{-d-\eta}\right)$ with some $\eta>0$ as $x\to\infty$ (condition of periodization).
\end{itemize}
\noindent
Introduce
\begin{itemize}
\item[--]
$\lambda_\beta\propto\sqrt{\beta}$ the thermal wave length,
\item[--]
$\rho=N/L^d$,
\item[--]
$\rho^{N,L}_n$ the density of particles in permutation cycles of length $n\geq 1$,
\item[--]
$\rho^{N,L}_0$ the density of zero-momentum particles,
\item[--]
$\rho_n=\lim_{N,L\to\infty,N/L^d=\rho}\,\rho^{N,L}_n$,  $n\geq 0$.
\end{itemize}
We have the following results.
\begin{enumerate}
\item
If $\int \hat{u}(x)x^2\d x<\infty$, then
\be\label{rho0-for-macr-cycles}
\rho_0=\lim_{\varepsilon\downarrow 0}\,\lim_{N,L\to\infty, N/L^d=\rho}\,\sum_{n\geq \varepsilon N}\rho^{N,L}_n .
\ee
\item
There exists a temperature-dependent positive number $\zeta_c(\beta)$ such that if $\rho>\zeta_c(\beta)/\lambda_\beta^d$ then
\be\label{infinite-cycles}
\sum_{n=1}^\infty \rho_n=\frac{\zeta_c(\beta)}{\lambda_\beta^d}
\quad\mbox{and}\quad
\lim_{\varepsilon\downarrow 0}\,\lim_{N,L\to\infty, N/L^d=\rho}\,\sum_{n\geq \varepsilon N}\rho^{N,L}_n
=\rho-\sum_{n=1}^\infty \rho_n>0.
\ee
\end{enumerate}
\end{theorem}
The best known example for a positive and positive-type interaction is a Gaussian function whose Fourier transform is also a Gaussian.
%(Substituting $v(x)=e^{-x^2}$ into (\ref{posdef}) yields $u(x)=(\pi/2)^{d/2} e^{-x^2/2}$.)
More generally, pair potentials that are both positive and positive type result as the autocorrelation function of some integrable nonnegative function $v$,
\be\label{posdef}
u(x)=\int_{\Rr^d} v(x+y)v(y)\d y,
\ee
in which case $\hat{u}(z)=|\hat{v}(z)|^2$.
The theorem is based on our previous work [Su11], [Su12] and on bounds for the free energy density inserted in the proof. In [Su11] we provided the tool by converting the Feynman-Kac formula for the partition function into a deterministic expression with the help of Fourier expansion. In [Su12] we applied the new formula to prove $\rho_0>0$ for positive-type pair potentials satisfying $\int \hat{u}(x)x^2\d x<\infty$ under the condition that there are cycles that together carry an asymptotically non-vanishing density and whose length diverges at least as fast as $N^{2/d}$. Our proof in [Su12] implied Eq.~(\ref{rho0-for-macr-cycles}) in the case when the possible infinite cycles are exclusively macroscopic, i.e. composed of a positive fraction of all the particles.
The new result is the inequality (\ref{infinite-cycles}). The sum $\sum_{n=1}^\infty \rho_n$ is the total density of particles in finite cycles in the infinite system, and the rest comes entirely from macroscopic cycles.
The upper and lower bounds on the free energy density will explain why we are able to show (\ref{infinite-cycles}) only if both $u$ and $\hat{u}$ are nonnegative: by
subtracting the mean field contribution from the potential energy the interaction loses its superstability but remains stable; meanwhile, the condition for BEC is unchanged. In this regard the system becomes similar to the noninteracting gas, and comparison with it facilitates the proof. The result is also similar. If any of $u\geq 0$ and $\hat{u}\geq 0$ fails, our proof fails as well.

A more fundamental reason why both  $u\geq 0$ and $\hat{u}\geq 0$ are necessary for the proof of (\ref{infinite-cycles}) is as follows.
 If $\hat{u}\geq 0$, at high densities the distribution of particles in classical ground states is uniform, while if $\hat{u}$ has a negative part, there is some structural order [Su5].
 Also, even if $\hat{u}\geq 0$ but $u$ is partly negative, at some intermediate densities the ground state can be ordered; this is the case, for example, when $\hat{u}$ is of compact support [Su13].
In general, if $u$ or $\hat{u}$ has a negative part, classical condensation or crystallization may take place at a higher temperature than that of the expected appearance of infinite cycles, so the latter should be proven in a restricted ensemble: in a liquid (the liquid helium) or in a crystal, cf. [Uel1].

Infinite cycles, off-diagonal long-range order (ODLRO) and BEC are three distinct notions. By definition, BEC implies ODLRO which implies infinite cycles, but implication in the opposite direction is subjected to conditions.
Both ODLRO and BEC are related to the infinite-volume limit of the integral kernel
$\langle x|\sigma^{N,L}_1| 0\rangle$ of the one-body reduced density matrix, cf. [Su12]. There is ODLRO if
\[
\lim_{x\to\infty}\,\lim_{N,L\to\infty, N/L^d=\rho}\langle x|\sigma^{N,L}_1|0\rangle\neq 0,
\]
including the possibility that the limit does not exist. The average of $\langle x|\sigma^{N,L}_1|0\rangle$ on the torus $\Lambda$ is the condensate density in $\Lambda$, and there is BEC if its limit is positive,
\[
\rho_0=\lim_{N,L\to\infty, N/L^d=\rho}\frac{1}{L^d}\int_\Lambda \langle x|\sigma^{N,L}_1| 0\rangle \d x> 0.
\]
Thus, in principle, ODLRO can exist without BEC, but the opposite is obviously false. Also, infinite cycles do not necessarily mean ODLRO. In [Su12] we found two conditions for ODLRO: (i) there must be cycles in the system whose length diverges with $N$ and which together carry a macroscopic number of particles, and (ii) given such a cycle of length $n$, the average shift, due to interaction, of the momentum of the particles forming the cycle
($h\overline{X^0_{^\cdot}}$, see in the proof) must go to zero as $n$ goes to infinity. According to [Su12], the additional condition for BEC is that $n$ diverges at least as fast as $N^{2/d}$, and $\overline{X^0_{^\cdot}}$ decays at least as fast as $1/\sqrt{n}$. The latter was shown to hold true if $\hat{u}\geq 0$, and
the proof below reveals that the density, carried by cycles whose length diverges but slower than $N$, tends to zero as $N$ goes to infinity.
The conclusion is that for pair potentials that are both positive and positive type the three notions coincide, and in the domain of BEC the cycles are either finite or macroscopic, just as in the noninteracting gas. Curiously, the inherent condition for BEC is weaker, the existence of cycles whose length diverges {\em at least} as fast as $N^{d/2}$.
The significance of $n\propto N^{2/d}\propto L^2/\lambda_\beta^2$ is clear: this is the order of magnitude of the necessary number of steps for a random walk of step length $\lambda_\beta$ that starts from zero to attain any point of a cube of side $L$.

It is worth recalling how the conditions on the pair potential strengthened from the first paper to the actual one. For the Fourier expansion of the Feynman-Kac formula the existence and integrability of $\hat{u}$ sufficed, not even stability was demanded. To prove that infinite cycles and ODLRO are simultaneous $\hat{u}\geq 0$ had to be supposed. Finally, to show in this paper that infinite cycles of the requested property do appear at high densities we need also $u\geq 0$. This does not mean that ODLRO and BEC are limited to such interactions, only a further extension calls for new ideas.

Section 2 contains the proof of the inequality (\ref{infinite-cycles}). The physical meaning of the permutation cycles is not obvious, a possible interpretation is given in Section 3. A survey of the long history of the research on BEC is presented in Section 4.

\newsec{Proof of the Theorem}

We start by writing the partition function in the form
\be\label{QNL-general}
Q_{N,L}=\frac{1}{N}\sum_{n=1}^N G^N_n
\ee
and defining $\rho^{N,L}_n$ by the equation
\be\label{rho_n/rho-general}
\frac{\rho^{N,L}_n}{\rho}=\frac{G^N_n}{N Q_{N,L}}.
\ee
The complete form of $G^N_n$ was given in [Su12] and will be recalled below. As explained in [Su12], $n$ is the length of a specific cycle, that one containing particle number 1. The particles are indistinguishable, whence the interpretation of $\rho^{N,L}_n$ as the density (number per unit volume) of particles in cycles of length $n$.
Because $\sum_{n=1}^N \rho^{N,L}_n= \rho$, with $\rho_n=\lim_{N,L\to\infty, N/L^d=\rho}\rho^{N,L}_n$ we have
\be\label{cut-at-M}
\rho=\lim_{M\to\infty}\lim_{N,L\to\infty, N/L^d=\rho}\left[\sum_{n=1}^M \rho^{N,L}_n
+ \sum_{M+1}^N  \rho^{N,L}_n\right]
= \sum_{n=1}^\infty\rho_n + \lim_{M\to\infty}\lim_{N,L\to\infty, N/L^d=\rho}\sum_{M+1}^N  \rho^{N,L}_n
\geq \sum_{n=1}^\infty\rho_n.
\ee
The infinite sum is the density of particles in finite cycles in the infinite system. The task is to show that if $\rho$ is larger than some $\beta$-dependent threshold value, the inequality is strict, so the double limit of the sum from $M+1$ to $N$ must be positive, signalling the presence of infinite cycles.

\vspace{5pt}
\noindent I.
Before dealing with interacting particles we briefly return to the ideal gas discussed in detail in [Su2] and revisited in [Su12].
The partition function is defined recursively by the equation
\be\label{Q0NL}
Q^0_{N,L}=\frac{1}{N}\sum_{n=1}^Nq_nQ^0_{N-n,L}\qquad (Q^0_{0,L}:=1)
\ee
where
\be\label{qn}
q_n=\sum_{z\in\Zz^d}\exp\left\{-\frac{\pi n \lambda_\beta^2}{L^2}z^2\right\}=\frac{L^d}{n^{d/2}\lambda_\beta^d}\sum_{z\in\Zz^d}\exp\left\{-\frac{\pi L^2}{n\lambda_\beta^2}z^2\right\}
\ee
is the partition function for a single particle at inverse temperature $n\beta$ ($n\lambda_\beta^2=\lambda_{n\beta}^2$).
Studying the identity
\be\label{1=undivided}
1=\frac{1}{N}\sum_{n=1}^Nq_n\frac{Q^0_{N-n,L}}{Q^0_{N,L}}
\ee
in the thermodynamic limit one can prove the appearance of infinite permutation cycles and their being exclusively macroscopic. In [Su2] this was achieved by detailed estimates concerning the symmetric group. In the alternative proof given in [Su12] the second form of $q_n$ was cut into the $z=0$ term and the rest, yielding
\be\label{unity-ideal}
1=\frac{1}{\rho\lambda_\beta^d}\sum_{n=1}^N\frac{1}{n^{d/2}}\frac{Q^0_{N-n,L}}{Q^0_{N,L}}
+\frac{1}{\rho\lambda_\beta^d}\sum_{n=1}^N\frac{1}{n^{d/2}}\frac{Q^0_{N-n,L}}{Q^0_{N,L}}\sum_{z\in\Zz^d\setminus\{0\}}\exp\left\{-\frac{\pi L^2}{n\lambda_\beta^2}z^2\right\}.
\ee
Using $Q^0_{N-n,L}/Q^0_{N,L}<1$ that follows from Eq.~(\ref{Q0NL}) and $q_n>1$, in Proposition 3.3 of [Su12] we then proved that the second sum over $n$ restricted to $n\leq K_N=o(N)$ tends to zero as $N,L$ go to infinity with $N/L^d=\rho$ fixed. On the other hand, the first sum saturates at $\zeta(d/2)$ where $\zeta$ is the Riemann zeta-function. So if $\zeta(d/2)/\rho\lambda_\beta^d<1$, the completion to 1 may come only from cycles of length $n\propto N$.
Here we give a variant of this proof, easier to extend to interacting particles.

Note first that with $N$ and $L$ increasing $Q^0_{N,L}=\exp\{-\beta F^0_{N,L}\}=\exp\{-\beta L^d [f^0(\rho,\beta)+o(1)]\}$ where $F^0_{N,L}$ is the free energy and $f^0(\rho,\beta)$ is the limiting free energy density of the ideal Bose gas. Furthermore, if $n=o(N)$ then
\be
\frac{Q^0_{N-n,L}}{Q^0_{N,L}}=e^{\beta n\frac{F^0_{N,L}/L^d-F^0_{N-n,L}/L^d}{(n/L^d)}}
=
e^{\beta n[\partial f^0(\rho,\beta)/\partial\rho+o(1)]}.
\ee
In $\partial f^0(\rho,\beta)/\partial\rho$ we recognise the chemical potential computed in the canonical ensemble.
Let $g_N=o(N^{2/d})$ be any monotone increasing sequence of integers that tends to infinity.
Applying the second form of $q_n$,
\be
\frac{1}{N}\sum_{n=1}^{g_N}q_n \frac{Q^0_{N-n,L}}{Q^0_{N,L}}
=\frac{1}{\rho\lambda_\beta^d}\sum_{n=1}^{g_N}\frac{1}{n^{d/2}}\left[1+O\left(e^{-\pi L^2/(n\lambda_\beta^2)}\right)\right]
e^{\beta n[\partial f^0(\rho,\beta)/\partial\rho+o(1)]}.
\ee
Proceeding as in Eq.~(\ref{cut-at-M}),
\bea\label{lim1}
\lim_{N,L\to\infty, N/L^d=\rho}\frac{1}{N}\sum_{n=1}^{g_N}q_n \frac{Q^0_{N-n,L}}{Q^0_{N,L}}
&=&
\frac{1}{\rho\lambda_\beta^d}\sum_{n=1}^\infty \frac{\exp\{\beta n\partial f^0(\rho,\beta)/\partial\rho\}}{n^{d/2}}
\nonumber\\
&+&
\lim_{M\to\infty}\lim_{N,L\to\infty, N/L^d=\rho}\frac{1}{N}\sum_{n=M+1}^{g_N}q_n \frac{Q^0_{N-n,L}}{Q^0_{N,L}}.
\eea
The infinite sum contains the contribution of all the finite cycles. The rest comes from cycles whose length although diverges with $N$, but the divergence is slower than $N^{2/d}$, and the double limit yields zero:
\[
\lim_{M\to\infty}\lim_{N,L\to\infty, N/L^d\rho}\frac{1}{N}\sum_{n=M+1}^{g_N}q_n \frac{Q^0_{N-n,L}}{Q^0_{N,L}}
=\frac{1}{\rho\lambda_\beta^d}\lim_{M\to\infty}\lim_{N,L\to\infty, N/L^d\rho}\sum_{n=M+1}^{g_N}\frac{1}{n^{d/2}} \frac{Q^0_{N-n,L}}{Q^0_{N,L}}=0.
\]
Next, fix any $c>0$ and choose any monotone increasing sequence $h_N$ of integers such that $h_N>c N^{2/d}$ and $h_N=o(N)$. Then
\be\label{lim2}
\lim_{N,L\to\infty, N/L^d=\rho}\frac{1}{N}\sum_{n=c N^{2/d}}^{h_N}q_n \frac{Q^0_{N-n,L}}{Q^0_{N,L}}
=0
\ee
because $q_n Q^0_{N-n,L}/Q^0_{N,L}=O(1)$. Combining Eqs.~(\ref{lim1}) and (\ref{lim2}),
\be\label{1=limit0}
1=\frac{1}{\rho\lambda_\beta^d}\sum_{n=1}^\infty \frac{\exp\{\beta n\partial f^0(\rho,\beta)/\partial\rho\}}{n^{d/2}}+
\lim_{\varepsilon\downarrow 0}\lim_{N,L\to\infty, N/L^d=\rho}\frac{1}{N}\sum_{n=\varepsilon N}^N\frac{Q^0_{N-n,L}}{Q^0_{N,L}}.
\ee
Here we used also that $q_n\to 1$ if $n$ increases faster than $N^{2/d}$.
Now $f^0$ is a negative strictly monotone decreasing function of $\rho$ below the critical density
$\zeta(d/2)/\lambda_\beta^d$. Therefore, for $\rho\leq\zeta(d/2)/\lambda_\beta^d$, $\varepsilon>0$ and $n\geq \varepsilon N$
\be
\frac{Q^0_{N-n,L}}{Q^0_{N,L}}\sim e^{-\beta L^d\left[f^0((1-n/N)\rho,\beta)-f^0(\rho,\beta)\right]}
\leq e^{-\beta L^d\left[f^0((1-\varepsilon)\rho,\beta)-f^0(\rho,\beta)\right]}
\ee
decays exponentially fast with the volume, $N^{-1}\sum_{n=\varepsilon N}^N Q^0_{N-n,L}/Q^0_{N,L}$ also decays at this rate and turns Eq.~(\ref{1=limit0}) into
\be
\frac{1}{\rho\lambda_\beta^d}\sum_{n=1}^\infty \frac{\exp\{\beta n\partial f^0(\rho,\beta)/\partial\rho\}}{n^{d/2}}=1\qquad (\rho\leq\zeta(d/2)/\lambda_\beta^d).
\ee
This equation determines $\partial f^0(\rho,\beta)/\partial\rho$ and assigns to it a non-positive value.
At $\rho=\zeta(d/2)/\lambda_\beta^d$ the sum over $n$ and $\partial f^0(\rho,\beta)/\partial\rho$ reach their maximum, $\zeta(d/2)$ and zero, respectively.
Therefore, if $\rho\lambda_\beta^d>\zeta(d/2)$, Eq.~(\ref{1=limit0}) can hold only with a positive contribution from macroscopic cycles. For this conclusion we are not obliged to further analyse the second term on the right-hand side of Eq.~(\ref{1=limit0}). Still, it must be detached from zero exactly when the first term sinks below 1, and this is easy to see.
If $\rho>\zeta(d/2)/\lambda_\beta^d$ then $\partial f^0(\rho,\beta)/\partial\rho\equiv 0$. At the same time, if
$(1-n/N)\rho\geq \zeta(d/2)/\lambda_\beta^d$, then
$f^0((1-n/N)\rho,\beta)=f^0(\rho,\beta)=-\frac{\zeta(1+d/2)}{\beta\lambda_\beta^d}$,
and $Q^0_{N-n,L}/Q^0_{N,L}\to 1$.
(In the case of periodic boundary conditions there is no surface/edge/corner correction to the limit of $Q^0_{N-n,L}/Q^0_{N,L}$.) So (\ref{1=limit0}) becomes
\be
\frac{\zeta(d/2)}{\rho\lambda_\beta^d}+\lim_{\varepsilon\downarrow 0}\lim_{N,L\to\infty, N/L^d=\rho}\frac{1}{N}\sum_{n=\varepsilon N}^{(1-\zeta(d/2)/\rho\lambda_\beta^d)N}\frac{Q^0_{N-n,L}}{Q^0_{N,L}}
=\frac{\zeta(d/2)}{\rho\lambda_\beta^d}+\left(1-\frac{\zeta(d/2)}{\rho\lambda_\beta^d}\right)=1\qquad (\rho>\zeta(d/2)/\lambda_\beta^d).
\ee

\vspace{5pt}
\noindent
II. For the partition function of the interacting Bose gas we will need both the Feynman-Kac formula and its Fourier-expanded form. Let $W^{n_l\beta}_{x_lx_l}(\d\omega_l)$ denote the Brownian bridge measure on the torus $\Lambda$ for trajectories $\omega_l$ that start and end in $x_l$ in the time interval $[0,n_l\beta]$. Define
\be\label{U(omega)}
U(\omega_l)=\beta^{-1}\int_0^\beta \sum_{0\leq j<k\leq n_l-1} u_L(\omega_l(k\beta+t)-\omega_l(j\beta+t))\d t,  \qquad l=0,\dots,p,
\ee
and
\be\label{U(omega,omega)}
U(\omega_{l'},\omega_l)=\beta^{-1}\int_0^\beta \sum_{j=0}^{n_{l'}-1}\sum_{k=0}^{n_l-1} u_L(\omega_l(k\beta+t)-\omega_{l'}(j\beta+t))\d t, \qquad
 0\leq l'<l\leq p
\ee
where
\be
u_L(x)=\sum_{z\in\Zz^d} u(x+Lz).
\ee
Then with $n_0=n$
\be
Q_{N,L}=\frac{1}{N}\sum_{n=1}^N \int_\Lambda\d x \int W^{n\beta}_{xx}(\d\omega_0)e^{-\beta U(\omega_0)} Q_{N-n,L}(\omega_0), \qquad N>0,
\ee
where $Q_{0,L}(\omega_0)=Q_{0,L}:=1$ and for $n<N$
\bea
Q_{N-n,L}(\omega_0)
&=&
\sum_{p=1}^{N-n}\frac{1}{p!}\sum_{\{n_l\}_{l=1}^p\vdash N-n}\frac{1}{\prod_{l=1}^pn_l}
\int_\Lambda \d x_1 \int W_{x_1x_1}^{n_1\beta}(\d\omega_1) e^{-\beta U(\omega_1)}
\cdots
\nonumber\\
&\cdots&
\int_\Lambda \d x_p \int W_{x_px_p}^{n_p\beta}(\d\omega_p) e^{-\beta U(\omega_p)}
\left(\prod_{1\leq l'<l\leq p}e^{-\beta  U(\omega_{l'},\omega_l)}\right)\,\,
e^{-\beta \sum_{l=1}^p U(\omega_0,\omega_l)}.
\eea
Without the last exponential that connects $\omega_0$ to the other cycles the above expression is just another form of $Q_{N-n,L}$.
Here and hereafter $\sum_{\{n_l\}_{l=1}^p\vdash N-n}\equiv \sum_{n_1,\dots,n_p\geq 1:\sum_1^p n_l=N-n}$, the same numbers in different order represent different terms. For the different ways to write the partition function as a multiple sum over the cycle lengths of permutations see [Su11].

We start by deriving bounds on the free energy $F_{N,L}$ and the free energy density
\be
f(\rho,\beta)
=\lim_{N,L\to\infty, N/L^d=\rho}L^{-d} F_{N,L}
=- \lim_{N,L\to\infty, N/L^d=\rho}\ \frac{1}{\beta L^d}\ln Q_{N,L}
\ee
for integrable superstable pair potentials [R2, R3].
\begin{lemma}
\be\label{FNL-upper-lower}
\frac{C_\Lambda[u]}{L^d}N(N-1)-BN+F^0_{N,L}
\leq F_{N,L}
\leq \frac{\|u\|_1}{2L^d}N(N-1)+\frac{2^{d/2-1}\zeta(d/2)\|u\|_1}{\lambda_\beta^d}N+F^0_{N,L}
\ee
and
\be\label{f-upper-lower}
C[u] \rho^2 - B\rho + f^0(\rho,\beta)
\leq f(\rho,\beta)
\leq
\frac{\|u\|_1}{2}\rho^2 + \frac{2^{d/2-1}\zeta(d/2)\|u\|_1}{\lambda_\beta^d}\rho + f^0(\rho,\beta)
\ee
where $B>0$ and $0<C_\Lambda[u],C[u]\leq \hat{u}(0)/2$.
If $u$ is bounded, for $B$ one can substitute $u_L(0)/2$ in the first line and $u(0)/2$ in the second line. For a positive-type $u$, $C_\Lambda[u]=C[u]=\hat{u}(0)/2$.
\end{lemma}
\noindent{\em Proof.} The lower bound is trivial except for the multipliers of $\rho$ and $\rho^2$. It follows from superstability: there are positive constants $B$, $C$ such that for $L$ sufficiently large and any $N$
\[
\sum_{1\leq j<k\leq N}u_L(x_k-x_j)\geq -BN+CN(N-1)/L^d.
\]
The largest $C$ is
\be\label{C_Lambda[u]}
C_\Lambda[u]=\liminf_{N\to\infty}\frac{L^d}{N(N-1)} \sum_{1\leq j<k\leq N}u_L(x_k-x_j)
\ee
that we called in [Su5] the best superstability constant (the Fekete constant in potential theory [Ch]). In general $C_\Lambda[u]\leq \hat{u}(0)/2$, but in Section 9 of [Su5] it was shown that for $u$ of the positive type $C_\Lambda[u]=\hat{u}(0)/2$ independently of $L$.
Even if $u\geq 0$, $B>0$ must be chosen, otherwise the inequality could fail for small $N$, e.g. if $u$ is of finite range. Also, for $N$ large the quadratic term alone with $C_\Lambda[u]$ can be too large. If $u$ is bounded, as implied by $\hat{u}\in L^1(\Rr^d)$,  $C=C_\Lambda[u]$ and $B=u_L(0)/2$ provide a valid lower bound.

For any partition of $N$
\[
\sum_{l=0}^p U(\omega_l)+\sum_{0\leq l'<l\leq p}U(\omega_{l'},\omega_l)
\geq - BN+C_\Lambda[u]  N(N-1)/L^d,
\]
hence
\begin{eqnarray*}
Q_{N,L}
&\leq& e^{-\beta\left[- BN+C_\Lambda[u] N(N-1)/L^d\right]}
\frac{1}{N}\sum_{n=1}^N \int_\Lambda\d x \int W^{n\beta}_{xx}(\d\omega_0)
\sum_{p=1}^{N-n}\frac{1}{p!}\sum_{\{n_l\}_{l=1}^p\vdash N-n}\prod_{l=1}^p\frac{1}{n_l} \int_\Lambda\d x_l \int W^{n_l\beta}_{x_lx_l}(\d\omega_l)
\nonumber\\
&=& e^{-\beta\left[- BN+C_\Lambda[u]  N(N-1)/L^d\right]} Q^0_{N,L}
\end{eqnarray*}
from which the lower bound for $f(\rho,\beta)$ follows with $u_L\to u$ as $L\to\infty$ and
\be\label{C[u]}
C[u]=\liminf_{L\to\infty}C_\Lambda[u].
\ee

The upper bound in (\ref{FNL-upper-lower}) and (\ref{f-upper-lower}) is based on Jensen's inequality. First, on the torus $Q_{N-n,L}(\omega_0+y)=Q_{N-n,L}(\omega_0)$, because any shift of $\omega_0$ can be defused by the same shift of the integration variables $x_1,\dots,x_p$ and hence of $\omega_1,\dots,\omega_p$; therefore
\[
Q_{N-n,L}(\omega_0)=\frac{1}{L^d}\int_\Lambda Q_{N-n,L}(\omega_0+y) \d y.
\]
The dependence on $\omega_0$ is only in the last exponential factor of $Q_{N-n,L}(\omega_0)$. Applying Jensen's inequality and
\[
\int_\Lambda u_L(y)\d y=\int u(y)\d y \leq \|u\|_1,
\]
\[
\frac{1}{L^d}\int_\Lambda e^{-\beta \sum_{l=1}^p U(\omega_0+y,\omega_l)} \d y
\geq e^{-\beta \frac{1}{L^d}\int_\Lambda \sum_{l=1}^p U(\omega_0+y,\omega_l) \d y}
= e^{-\beta \frac{n(N-n)}{L^d}\int_\Lambda u_L(y)\d y}
\geq e^{-\beta \frac{n(N-n)}{L^d} \|u\|_1}.
\]
Thus, for any $\omega_0$
\be
Q_{N-n,L}(\omega_0)\geq e^{-\beta \frac{n(N-n)}{L^d} \|u\|_1} Q_{N-n,L}
\ee
and consequently
\be\label{QNL-lower}
Q_{N,L}\geq \frac{1}{N}\sum_{n=1}^N  \exp\left\{-\beta \|u\|_1\frac{n(N-n)}{L^d} \right\}  Q_{N-n,L}\,
\int_\Lambda\d x \int W^{n\beta}_{xx}(\d\omega_0)e^{-\beta U(\omega_0)}.
\ee
Recall from [Su11] that
\[
\int W^{n\beta}_{xx}(\d\omega)=\sum_{z\in\Zz^d}\int P^{n\beta}_{x,x+Lz}(\d\omega)
= \frac{1}{\lambda_{n\beta}^d}\sum_{z\in\Zz^d}e^{-\pi L^2 z^2/\lambda_{n\beta}^2},
\]
therefore
\[
\int_\Lambda \d x \int W^{n\beta}_{xx}(\d\omega)
=\frac{L^d}{\lambda_{n\beta}^d}\sum_{z\in\Zz^d}e^{-\pi L^2 z^2/\lambda_{n\beta}^2}=q_n.
\]
Above $P^{n\beta}_{x,x+Lz}(\d\omega)$ is the conditional Wiener measure in $\Rr^d$. Let
$
\psi_t(x)=\lambda_t^{-d}e^{-\pi x^2/\lambda_t^2},
$
then
\[\label{norms}
\int P^\beta_{xy}(\d\omega)
=\psi_\beta(y-x).
\]
Throughout the proof we use the following identity.
If $f(\omega)=f(\omega(t_0))$ with $0<t_0<\beta$, then
\be\label{one-point}
\int P^\beta_{xy}(\d\omega)f(\omega)=\int\d x'\psi_{t_0}(x-x')f(x')\psi_{\beta-t_0}(x'-y).
\ee
By Jensen's inequality,
\bea\label{Jensen}
\int_\Lambda \d x \int W^{n\beta}_{xx}(\d\omega) e^{-\beta U(\omega)} = L^d\int W^{n\beta}_{00}(\d\omega) e^{-\beta U(\omega)}
=L^d \sum_{z\in\Zz^d}\int P^{n\beta}_{0,Lz}(\d\omega) e^{-\beta U(\omega)}
\nonumber\\
\geq
L^d\sum_{z\in\Zz^d}\left(\int P^{n\beta}_{0,Lz}(\d\omega)\right)\exp\left\{-\beta
\frac{\int P^{n\beta}_{0,Lz}(\d\omega)U(\omega)}{ \int P^{n\beta}_{0,Lz}(\d\omega)}\right\}
\nonumber\\
=
\frac{L^d}{n^{d/2}\lambda_{\beta}^d}\sum_{z\in\Zz^d}\exp\left\{-\frac{\pi L^2 z^2}{n\lambda_\beta^2}\right\}
\exp\left\{-\beta\langle U\rangle_{P^{n\beta}_{0,Lz}}\right\}.
\eea
We turn to $\langle U\rangle_{P^{n\beta}_{0,Lz}}$.
With a slight extension of (\ref{one-point})
one can show that equal-time increments have the same distribution. Let $0<t_1<t_2<\beta$, $\omega(0)=x$, $\omega(\beta)=y$, and consider any $f$ depending only on $\omega(t_2)-\omega(t_1)$. Then
\begin{eqnarray*}
\int P^\beta_{xy}(\d\omega)f(\omega(t_2)-\omega(t_1))
&=&\int\d x_1\d x_2\ \psi_{t_1}(x_1-x)\psi_{t_2-t_1}(x_2-x_1)\psi_{\beta-t_2}(y-x_2)f(x_2-x_1)
\nonumber\\
&=&\int\d z\ \psi_{t_2-t_1}(z)f(z)\int\d x_1\ \psi_{t_1}(x_1-x)\psi_{\beta-t_2}(y-z-x_1)
\nonumber\\
&=&\int\d z\ \psi_{t_2-t_1}(z)f(z)\psi_{\beta-(t_2-t_1)}(y-x-z)
=\int P^\beta_{0,y-x}(\d\omega)f(\omega(t_2-t_1)).
\end{eqnarray*}
Thus,
\[
\int P^{n\beta}_{0,Lz}(\d\omega) u_L(\omega(k\beta+t)-\omega(j\beta+t))
=\int P^{n\beta}_{0,Lz}(\d\omega) u_L(\omega((k-j)\beta))
\]
and
\be\label{U-average-two-forms}
\langle U\rangle_{P^{n\beta}_{0,Lz}}
=\sum_{0\leq j<k\leq n-1}\langle u_L(\omega((k-j)\beta))\rangle_{P^{n\beta}_{0,Lz}}
=\sum_{k=1}^{n-1}(n-k)\langle u_L(\omega(k\beta))\rangle_{P^{n\beta}_{0,Lz}}
=\sum_{k=1}^{n-1}k\langle u_L(\omega((n-k)\beta))\rangle_{P^{n\beta}_{0,Lz}}.
\ee
For any $z\in\Zz^d$,
$
u_L(x)=u_L(Lz-x),
$
from which with (\ref{one-point}) one obtains
\[
 \int P^{n\beta}_{0,Lz}(\d\omega) u_L(\omega((n-k)\beta))
=\int P^{n\beta}_{0,Lz}(\d\omega)  u_L(\omega(k\beta)).
\]
Summing the two forms (\ref{U-average-two-forms}) of $\langle U\rangle_{P^{n\beta}_{0,Lz}}$ and using the above equality,
\be\label{sum-of-two-forms}
\langle U\rangle_{P^{n\beta}_{0,Lz}}=\frac{n}{2}\sum_{k=1}^{n-1}\langle u_L(\omega(k\beta))\rangle_{P^{n\beta}_{0,Lz}}.
\ee
Next,
%we evaluate $\langle u_L(\omega(k\beta))\rangle_{P^{n\beta}_{0,Lz}}$.
\begin{eqnarray*}
\langle u_L(\omega(k\beta))\rangle_{P^{n\beta}_{0,Lz}}
&=&\left(\int P^{n\beta}_{0,Lz}(\d\omega) \right)^{-1}\int P^{n\beta}_{0,Lz}(\d\omega) u_L(\omega(k\beta))
\nonumber\\
&=&\psi_{n\beta}(Lz)^{-1}\int  \psi_{k\beta}(x)\psi_{(n-k)\beta}(Lz-x) u_L(x) \d x.
\end{eqnarray*}
A straightforward calculation yields
\[
\psi_{n\beta}(Lz)^{-1} \psi_{k\beta}(x)\psi_{(n-k)\beta}(Lz-x)
=\alpha_{n,k}^{d/2}\exp\{-\pi\alpha_{n,k}(x-Lkz/n)^2\}
\]
and therefore
\be\label{av-u_L}
\langle u_L(\omega(k\beta))\rangle_{P^{n\beta}_{0,Lz}}
=\alpha_{n,k}^{d/2}\int u_L(x)\exp\left\{-\pi\alpha_{n,k}(x-Lkz/n)^2\right\}\d x
\ee
where
\[
\alpha_{n,k}=\left(\frac{1}{k}+\frac{1}{n-k}\right)\frac{1}{\lambda_\beta^2}.
\]
Substituting $u_L(x)=\sum_{v\in\Zz^d}u(x-Lv)$ into (\ref{av-u_L}),
\begin{eqnarray*}
\langle u_L(\omega(k\beta))\rangle_{P^{n\beta}_{0,Lz}}
&=&
\alpha_{n,k}^{d/2}\sum_{v\in\Zz^d}\int u(x-Lv)\exp\left\{-\pi\alpha_{n,k}(x-Lkz/n)^2\right\}\d x
\nonumber\\
&=&
\int u(y)\ \alpha_{n,k}^{d/2}\sum_{v\in\Zz^d}\exp\left\{-\pi\alpha_{n,k}(y-L\{kz/n\}+Lv)^2\right\}\d y
\end{eqnarray*}
where $\{kz/n\}$ is the fractional part of $kz/n$ for $z\in\Zz^d$, each component of which is bounded in modulus by $1/2$.
Applying twice the Poisson summation formula
\begin{eqnarray*}
\alpha_{n,k}^{d/2}\sum_{v\in\Zz^d}\exp\left\{-\pi\alpha_{n,k}(y-L\{kz/n\}+Lv)^2\right\}
=L^{-d}\sum_{v\in\Zz^d}\exp\left\{-\frac{\pi v^2}{\alpha_{n,k}L^2}+\i \frac{2\pi}{L}(y-L\{kz/n\})\cdot v\right\}
\nonumber\\
\leq L^{-d}\sum_{v\in\Zz^d}\exp\left\{-\frac{\pi v^2}{\alpha_{n,k}L^2}\right\}
=\alpha_{n,k}^{d/2}\sum_{v\in\Zz^d}\exp\left\{-\pi\alpha_{n,k}L^2v^2\right\}.
\end{eqnarray*}
So
\be
\langle u_L(\omega(k\beta))\rangle_{P^{n\beta}_{0,Lz}}
\leq \|u\|_1\alpha_{n,k}^{d/2}\sum_{v\in\Zz^d}\exp\left\{-\pi\alpha_{n,k}L^2v^2\right\},
\ee
the $z$-dependence dropped from the upper bound.
Now
\begin{eqnarray*}
\sum_{v\in\Zz^d}e^{-\pi\alpha_{n,k}L^2v^2}=\left(1+2\sum_{v=1}^\infty e^{-\pi\alpha_{n,k}L^2v^2}\right)^d
\leq \left(1+2\int_0^\infty e^{-\pi\alpha_{n,k}L^2v^2}\d v  \right)^d
=\left(1+\frac{1}{L\sqrt{\alpha_{n,k}}}\right)^d.
\end{eqnarray*}
Inserting this into the expression for $\langle U\rangle_{P^{n\beta}_{0,Lz}}$,
\begin{eqnarray*}
\langle U\rangle_{P^{n\beta}_{0,Lz}}
&\leq&\frac{n}{2} \|u\|_1\sum_{k=1}^{n-1}\alpha_{n,k}^{d/2}\sum_{v\in\Zz^d}\exp\left\{-\pi\alpha_{n,k}L^2v^2\right\}
\leq \frac{n}{2} \|u\|_1\sum_{k=1}^{n-1}\alpha_{n,k}^{d/2}\left(1+\frac{1}{L\sqrt{\alpha_{n,k}}}\right)^d
\nonumber\\
&=&\frac{n}{2} \|u\|_1\sum_{l=0}^d {d\choose l}\frac{1}{L^{d-l}}\sum_{k=1}^{n-1}\alpha_{n,k}^{l/2}
=\frac{n}{2} \|u\|_1 \left[\frac{n-1}{L^d}+\sum_{l=1}^{d-1}{d\choose l}\frac{1}{L^{d-l}}\sum_{k=1}^{n-1}\alpha_{n,k}^{l/2}+\sum_{k=1}^{n-1}\alpha_{n,k}^{d/2}\right].
\end{eqnarray*}
For $d\geq 3$ and $1\leq l\leq d-1$
\[
\frac{1}{L^{d-l}}\sum_{k=1}^{n-1}\alpha_{n,k}^{l/2}
= \frac{1}{\lambda_\beta^l L^{d-l}}\sum_{k=1}^{n-1}\left(\frac{1}{k}+\frac{1}{n-k}\right)^{l/2}
\leq \frac{2^{l/2}}{\lambda_\beta^l L^{d-l}}\sum_{k=1}^{n/2}\frac{1}{k^{l/2}}=o(1)\quad (L\to\infty)
\]
which finally yields
\be\label{av-U-final-upper-bound}
\langle U\rangle_{P^{n\beta}_{0,Lz}}
\leq \frac{\|u\|_1}{2} n \left[\frac{n-1}{L^d}+\sum_{k=1}^{n-1}\alpha_{n,k}^{d/2}+o(1)  \right]
\leq \frac{\|u\|_1}{2} n \left[\frac{n-1}{L^d}+2^{d/2}\zeta(d/2)/\lambda_\beta^d +o(1)\right].
\ee
Dropping the $o(1)$ term,
the inequality (\ref{QNL-lower}) can be continued as
\bea
Q_{N,L}
&\geq&
\frac{1}{N}\sum_{n=1}^N q_n  Q_{N-n,L}
\exp\left\{- \frac{\beta\|u\|_1}{L^d} \left[n(N-n)+\frac{1}{2}n(n-1)\right]\right\}
\exp\left\{-2^{d/2-1}\zeta(d/2)\frac{\beta \|u\|_1}{\lambda_{\beta}^d}n\right\}
\nonumber\\
&\equiv&
\frac{1}{N}\sum_{n=1}^N q_n  Q_{N-n,L}
\exp\left\{- C \left[n(N-n)+\frac{1}{2}n(n-1)\right]\right\}
\exp\left\{-Dn\right\}
\equiv
\frac{1}{N}\sum_{n=1}^N q_n  Q_{N-n,L}\, e^{-\beta\Psi^+_{n,N-n}}.
\nonumber\\
\eea
We define an auxiliary function $Q_{N,L}^-$ recursively by $Q_{0,L}^-=1$ and
\be\label{QNL-minus}
Q_{N,L}^-=\frac{1}{N}\sum_{n=1}^N q_nQ_{N-n,L}^- e^{-\beta\Psi^+_{n,N-n}}.
\ee
$Q_{N,L}^-$ has two useful properties.

\vspace{3pt}
\noindent
(i) $Q_{N,L}\geq Q_{N,L}^-$. To see it, write $Q_{N,L}$ in an analogous form,
\be\label{QN-bis}
Q_{N,L}=\frac{1}{N}\sum_{n=1}^N q_nQ_{N-n,L}e^{-\beta\Psi_{n,N-n}}.
\ee
By definition
\[
e^{-\beta\Psi_{n,N-n}}=\frac{\int_\Lambda\d x \int W^{n\beta}_{xx}(\d\omega_0)e^{-\beta U(\omega_0)} Q_{N-n,L}(\omega_0)}{q_nQ_{N-n,L}},
\]
and we just have proved that $\Psi_{n,N-n}\leq\Psi^+_{n,N-n}$. Therefore
\bea
Q_{N,L}-Q_{N,L}^-
&=&\frac{1}{N}\sum_{n=1}^N q_n\left[Q_{N-n,L}e^{-\beta\Psi_{n,N-n}}-Q_{N-n,L}^-e^{-\beta\Psi^+_{n,N-n}}\right]
\nonumber\\
&=&\frac{1}{N}\sum_{n=1}^N q_n e^{-\beta\Psi_{n,N-n}}\left[Q_{N-n,L}-Q_{N-n,L}^-e^{-\beta(\Psi^+_{n,N-n}-\Psi_{n,N-n})}\right]
\nonumber\\
&\geq&\frac{1}{N}\sum_{n=1}^N q_n e^{-\beta\Psi_{n,N-n}}\left[Q_{N-n,L}-Q_{N-n,L}^-\right].
\eea
For $N=1$ this reads
\be
Q_{1,L}-Q_{1,L}^-\geq q_1 e^{-\beta\Psi_{1,0}}[Q_{0,L}-Q_{0,L}^-]=0,
\ee
and $Q_{N,L}\geq Q_{N,L}^-$ follows by induction.

\vspace{3pt}
\noindent
(ii)
\be\label{QNL-minus-vs-QNL0}
Q_{N,L}^-= \exp\left\{-\frac{1}{2}CN(N-1)-DN\right\}Q^0_{N,L}.
\ee
Indeed, apply the identity
\begin{eqnarray*}
\frac{1}{2}CN(N-1)+DN
&=&Cn(N-n) +\frac{1}{2}Cn(n-1)+Dn+\frac{1}{2}C(N-n)(N-n-1)+D(N-n)
\nonumber\\
&=&\beta\Psi^+_{n,N-n}+ \frac{1}{2}C(N-n)(N-n-1)+D(N-n).
\end{eqnarray*}
From the definition (\ref{QNL-minus}) one can see that
$A_N=e^{\frac{1}{2}CN(N-1)+DN}Q_{N,L}^-$ satisfies the recurrence relation
\be
A_N=\frac{1}{N}\sum_{n=1}^N q_n  A_{N-n}
\ee
with the initial condition $A_0=1$.
$Q^0_{N,L}$ satisfies the same equation with the same initial condition, cf. Eq.~(\ref{Q0NL}), therefore $A_N=Q^0_{N,L}$, which proves (\ref{QNL-minus-vs-QNL0}). In sum,
\be
Q_{N,L}
\geq \exp\left\{-\frac{1}{2}CN(N-1)-DN\right\}Q^0_{N,L}
=\exp\left\{-\frac{1}{2}\beta\rho\|u\|_1(N-1)
-2^{d/2-1}\zeta(d/2)\frac{\beta\|u\|_1}{\lambda_\beta^d}N\right\} Q^0_{N,L}.
\ee
From here the upper bound for $F_{N,L}$ and $f(\rho,\beta)$ follows. $\quad\Box$

\vspace{3pt}
It is instructive to replace $Q_{N,L}$ with $\tilde{Q}_{N,L}=e^{\beta\|u\|_1N(N-1)/2L^d}Q_{N,L}$. The observables like $\rho^{N,L}_{n}$ are invariant under this change;
the difference is in the free energy densities, $f(\rho,\beta)$ and
\be\label{logQ-tilde}
\tilde{f}(\rho,\beta)=- \lim_{N,L\to\infty, N/L^d=\rho}\ \frac{1}{\beta L^d}\ln \tilde{Q}_{N,L}=f(\rho,\beta) - \rho^2 \|u\|_1/2.
\ee
According to the Lemma, for $\tilde{f}(\rho,\beta)$ we have the bounds
\be\label{lower-upper}
\left(C[u]-\|u\|_1/2\right) \rho^2 - B\rho + f^0(\rho,\beta)
\leq \tilde{f}(\rho,\beta)
\leq
 \frac{2^{d/2-1}\zeta(d/2)\|u\|_1}{\lambda_\beta^d}\rho + f^0(\rho,\beta)
\ee
with $C[u]\leq \hat{u}(0)/2\leq\|u\|_1/2$.
In [Su5] we studied the $\rho\to\infty$ limit of $\rho^{-2}f(\rho,\infty)$ for classical systems, and found that it was bounded above by the best superstability constant, cf. Eqs.~(\ref{C_Lambda[u]}) and (\ref{C[u]}), which in turn cannot exceed $\hat{u}(0)/2$ and reaches this value for $u$ of the positive type. Thus, classically and without the kinetic energy (\ref{logQ-tilde}) is negative, while for integrable superstable potentials at high densities $f(\rho,\beta)$ is positive and increases as $\rho^2$.
If $Q_{N,L}$ is the partition function of the mean-field Bose gas with mean-field energy $(\|u\|_1/2)N(N-1)/L^d$
then $\tilde{Q}_{N,L}$ is the partition function of the ideal Bose gas. Its energy is purely kinetic, yet (\ref{logQ-tilde}) is negative due to the known particularity of the Bose statistics that it induces an effective attraction among the particles.
The result can be similar for a positive-type $u$ because of the uniform distribution of the particles [Su5] and also because at asymptotically high densities the ground state energy density of the Bose gas is $\rho^2\hat{u}(0)/2$, cf. Lieb [Li7].
This can also be seen on (\ref{lower-upper}) which for a positive and positive-type $u$, when $C[u]=\|u\|_1/2$, becomes
\be\label{bounds-on-tilde{f}}
- \frac{u(0)}{2}\rho + f^0(\rho,\beta)
\leq \tilde{f}(\rho,\beta)
\leq
 \frac{2^{d/2-1}\zeta(d/2)\hat{u}(0)}{\lambda_\beta^d}\rho + f^0(\rho,\beta),
\ee
the correction to $\rho^2\hat{u}(0)/2$ is of order $\rho$. The upper bound in (\ref{lower-upper}) goes to zero with the temperature, but not fast enough to decide whether $\tilde{f}(\rho,\beta)$ could be negative (as $\beta$ increases, $\rho$ exceeds the critical density $\zeta(d/2)/\lambda_\beta^d$ of the ideal gas, above which $f^0(\rho,\beta)=-\zeta(1+d/2)/(\beta\lambda_\beta^d)$ tends to zero faster than the positive first term). However, the increase, if any, of $\tilde{f}(\rho,\beta)$ with $\rho$ is at most linear and is probably due to the increase of the kinetic energy.
There is a sensible difference in the negative lower bound: it is linear in $\rho$ if both $u$ and $\hat{u}$ are nonnegative and quadratic if either $u$ or $\hat{u}$ is partly negative. In the first case, in analogy with the relation between the mean-field and the ideal Bose gas, the change from $Q_{N,L}$ to $\tilde{Q}_{N,L}$ deprives the model from its superstability but preserves stability; in the second case not only the superstability but possibly also the stability is lost.

The inequalities (\ref{bounds-on-tilde{f}}) give already a hint to the existence of BEC. They imply $\lim_{\rho\to\infty}f(\rho,\beta)/\rho^2=\hat{u}(0)/2$, extending the ground-state result to positive temperatures. This shows that at asymptotically high densities a positive and positive-type pair potential acts as a mean-field interaction. $f(\rho,\beta)$ is a convex function of $\rho$ and $\partial f(\rho,\beta)/\partial \rho$ is the chemical potential of the canonical ensemble.
The pair potential $\tilde{u}_L(x)=u_L(x)-\hat{u}(0)/L^d$ is still stable, therefore it defines a normal thermodynamic system, but $\tilde{f}(\rho,\beta)=f(\rho,\beta) - \rho^2 \hat{u}(0)/2$ may not be convex. If it is, the chemical potential $\partial \tilde{f}/\partial\rho$ is an increasing function of $\rho$. However, due to (\ref{bounds-on-tilde{f}}) it is bounded from above, and if it attains its supremum at a finite $\rho$ then for higher densities the surplus particles must go into the condensate. Below we have to circumvent the question of convexity of $\tilde{f}(\rho,\beta)$.

For interacting particles the Fourier-expanded form of the Feynman-Kac formula for the partition function reads
\be\label{QN}
Q_{N,L}=
\frac{e^{-\beta\hat{u}(0)N(N-1)/2L^d}}{N}
\sum_{n=1}^{N} \Phi^N_{n}\qquad (Q_{0,L}:=1).
\ee
We give below the asymptotic form of $\Phi^N_{n}$, valid for $L$ so large that the Riemann integral-approximating sum $L^{-d}\sum_{z\in\Zz^d\setminus\{0\}}\hat{u}(z/L)$ can be replaced with $\int_{\Rr^d} \hat{u}(x) \d x$, cf. [Su11, Su12].
For $n=N$
%%%%%%%%%%%%%%%%%%%%%%%%%%
\bea\label{PhiNN}
\Phi^N_N=\sum_{\{\alpha^k_j\in \Nn_0|1\leq j<k\leq N\}}
\ \prod_{1\leq j<k\leq N}
\frac{ \left(-\beta\right)^{\alpha^k_j}}{\alpha^k_j !}
\prod_{r=1}^{\alpha^k_j}\int_0^1\d t^k_{j,r} \int \d x^k_{j,r}\ \hat{u}\left(x^k_{j,r}\right)
\nonumber\\
\exp\left\{-\pi N \lambda_\beta^2\left[\overline{\left(X^0_{^\cdot}\right)^2}-\overline{X^0_{^\cdot}}^2\right]\right\}q_{N}(\overline{X^0_{^\cdot}})
\eea
%%%%%%%%%%%%%%%%%%%%%%%%%%%%%
and for $n<N$
%%%%%%%%%%%%%%%%%%%%%%%%%%%%
\be\label{Phi^N_{n_0}}
\Phi^N_{n}=\sum_{p=1}^{N-n}\frac{1}{p!}\sum_{n_1,\dots,n_p\geq 1:\sum_1^p n_l=N-n}\ \frac{1}{\prod_{l=1}^p n_l}\ \Phi^N_{n,\{n_l\}_1^p}
\ee
where ($n_0=n$)
\bea\label{Phi^N_{n,{n_l}_1^p}}
\Phi^N_{n,\{n_l\}_1^p}
=
\sum_{\{\alpha^k_j\in \Nn_0|1\leq j<k\leq N\}}
\Delta_{\{\alpha^k_j\},\{n_l\}_0^p}\, L^{-dK_{\{\alpha^k_j\}}}
 \prod_{1\leq j<k\leq N}
\frac{ \left(-\beta\right)^{\alpha^k_j}}{\alpha^k_j !}
\prod_{r=1}^{\alpha^k_j}\int_0^1\d t^k_{j,r} \int \d x^k_{j,r}\ \hat{u}\left(x^k_{j,r}\right)    %}
\nonumber\\
\left[
\delta(X^0_1,\dots,X^p_1)
\prod_{l=0}^p
\exp\left\{-\pi n_l \lambda_\beta^2\left[\overline{\left(X^l_{^\cdot}\right)^2}-\overline{X^l_{^\cdot}}^2\right]\right\}
q_{n_l}(\overline{X^l_{^\cdot}})\right].
\nonumber\\
\eea
The content in the outer square brackets is in the argument of the multiple integral and hence depends on all the summation and integration variables. To connect with the notations of the preceding papers, $\Phi^N_{n,\{n_l\}_1^p}$ agrees with $L^d F\left[n,\{n_l\}_1^p\right](0)=\exp\{\beta\hat{u}(0)N(N-1)/2L^d\}G\left[n,\{n_l\}_1^p\right]$, where $L^d F\left[n,\{n_l\}_1^p\right](x)$ played a central role in [Su12] and $G\left[n,\{n_l\}_1^p\right]$ is the Feynman-Kac formula which was our starting point in [Su11]. We repeat here the original expression using the definitions (\ref{U(omega)}), (\ref{U(omega,omega)}):
\bea\label{Phi^N_n-FK}
\Phi^N_{\{n_l\}_0^p}=e^{\beta\hat{u}(0)N(N-1)/2L^d}
\int_\Lambda \d x_0 \int W_{x_0x_0}^{n_0\beta}(\d\omega_0) e^{-\beta U(\omega_0)}\dots
\nonumber\\
\dots
\int_\Lambda \d x_p \int W_{x_px_p}^{n_p\beta}(\d\omega_p)
e^{-\beta U(\omega_p)}
\prod_{0\leq l'<l\leq p} e^{-\beta U(\omega_{l'},\omega_l)}.
\eea
Because
\[
\sum_{p=1}^{N-n}\frac{1}{p!}\sum_{\{n_l\}_{l=1}^p\vdash N-n}\prod_{l=1}^p\frac{1}{n_l}=1,
\]
$\Phi^N_n$ is the average of $\Phi^N_{n,\{n_l\}_1^p}$ over the partitions $\{n_l\}_1^p$ of $N-n$.
%%%%%%%%%%%%%%%%%%%%%%%%%%%%%%%%%%%%
The entries of $\Phi^N_{n}$ are defined as follows. Let
\bea
N_l
&=&
\sum_{l'=0}^l n_{l'}\quad (l=0,1,\dots,p),\quad N_0= n_0=n, \quad N_p=N,
\nonumber\\
C_l
&=&
\{N_{l-1}+1,\dots,N_l\},
\eea
the set of numbers labelling the particles of cycle $l$. The quantities denoted by $X$ are derived from $X_q(t)=Z_q(t)/L$, and the definition of $Z_q(t)$ is at the beginning of [Su11]. In $X_q(t)$ we have to separate intra- and inter-cycle  terms.
The two contributions are labelled by $l$ and $\neg l$, respectively. Hence, for $q\in C_l$
\bea\label{X-decomposition}
X_q(t)
&=&
X_q(t)|_l+X_q(t)|_{\neg l}
\nonumber\\
X_q(t)|_l
&=&
-\sum_{j=N_{l-1}+1}^{q-1}\sum_{k=q}^{N_l}\sum_{r=1}^{\alpha^{k}_{j}}{\bf 1}\{  t^{k}_{j,r}\geq t\} x^{k}_{j,r}
-\sum_{j=N_{l-1}+1}^{q}\sum_{k=q+1}^{N_l}\sum_{r=1}^{\alpha^{k}_{j}}{\bf 1}\{  t^{k}_{j,r}<t\} x^{k}_{j,r}
\nonumber\\
X_q(t)|_{\neg l}
&=&
-\sum_{j=1}^{N_{l-1}}\sum_{k=q}^{N_l}\sum_{r=1}^{\alpha^{k}_{j}}{\bf 1}\{  t^{k}_{j,r}\geq t\} x^{k}_{j,r}
-\sum_{j=1}^{N_{l-1}}\sum_{k=q+1}^{N_l}\sum_{r=1}^{\alpha^{k}_{j}}{\bf 1}\{  t^{k}_{j,r}<t\} x^{k}_{j,r}
\nonumber\\
&&
+\sum_{j=q}^{N_l}\sum_{k=N_l+1}^{N}\sum_{r=1}^{\alpha^{k}_{j}}{\bf 1}\{  t^{k}_{j,r}\geq t\} x^{k}_{j,r}
+\sum_{j=q+1}^{N_l}\sum_{k=N_l+1}^{N}\sum_{r=1}^{\alpha^{k}_{j}}{\bf 1}\{  t^{k}_{j,r}<t\} x^{k}_{j,r}.
\eea
Note that the quantities occurring in $\delta(X^0_1,\dots,X^p_1)$,
\be\label{X^l_1}
X^l_1 := X_{N_{l-1}+1}(0)
=-\sum_{j=1}^{N_{l-1}}\sum_{k\in C_l}\sum_{r=1}^{\alpha^{k}_{j}}x^{k}_{j,r}+\sum_{j\in C_l} \sum_{k=N_l+1}^N \sum_{r=1}^{\alpha^{k}_{j}}x^{k}_{j,r},
\quad l=0,\dots,p
\ee
are purely inter-cycle. Now
$\overline{X^l_{^\cdot}}$ and $\overline{\left(X^l_{^\cdot}\right)^2}$ are the averages of $X_q(t)$ and $X_q(t)^2$, respectively:
\bea\label{Xav-decomposition}
\overline{X^l_{^\cdot}}
&=&
\frac{1}{n_l}\sum_{q\in C_l}\int_0^1 X_q(t)|_l\,\d t+\frac{1}{n_l}\sum_{q\in C_l}\int_0^1 X_q(t)|_{\neg l}\,\d t
= \overline{X^l_{^\cdot}|_l} + \overline{X^l_{^\cdot}|_{\neg l}}
\nonumber\\
\overline{\left(X^l_{^\cdot}\right)^2}
&=&
\frac{1}{n_l}\sum_{q\in C_l}\int_0^1 \left[X_q(t)|_l+X_q(t)|_{\neg l}\right]^2\d t
\nonumber\\
&=&
\frac{1}{n_l}\sum_{q\in C_l}\int_0^1 \left(X_q(t)|_l\right)^2\d t + \frac{1}{n_l}\sum_{q\in C_l}\int_0^1\left(X_q(t)|_{\neg l}\right)^2\d t
+\frac{2}{n_l}\sum_{q\in C_l}\int_0^1 \left(X_q(t)|_l \cdot X_q(t)|_{\neg l}\right) \d t
\nonumber\\
&=&
\overline{\left(X^l_{^\cdot}\right)^2|_l }+ \overline{\left(X^l_{^\cdot}\right)^2|_{\neg l}} + 2\,\overline{X^l_{^\cdot}|_l\cdot X^l_{^\cdot}|_{\neg l}}.
\eea
The averages were computed in [Su11]. Referring to the expressions there, $x^k_{j,r}=z^k_{j,r}/L$, $\overline{X^l_{^\cdot}}=\overline{Z^l_{^\cdot}}/L$ and $\overline{(X^l_{^\cdot})^2}=\overline{(Z^l_{^\cdot})^2}/L^2$.
The only difference with [Su11] is that here, as in [Su12], $l$ goes from 0 to $p$ and $l=0$ is treated separately.

Furthermore, for $l=0,\dots,p$
%%%%%%%%%%%%%%%%%%%%%
\be\label{qnl}
q_{n_l}(\overline{X^l_{^\cdot}})=\sum_{z\in\Zz^d}\exp\left\{-\frac{\pi n_l \lambda_\beta^2}{L^2}\left(z+\left\{L\overline{X^l_{^\cdot}}\right\}\right)^2\right\}
=\frac{L^d}{n_l^{d/2}\lambda_{\beta}^d}\sum_{z\in\Zz^d}\exp\left\{-\frac{-\pi L^2z^2}{n_l\lambda_\beta^2}\right\}
\cos\left(2\pi \left\{L\overline{X^l_{^\cdot}}\right\} \cdot z\right).
\ee
Originally one has $L\overline{X^l_{^\cdot}}$ in this formula, but
decomposing it as $L\overline{X^l_{^\cdot}}=[L\overline{X^l_{^\cdot}}]+\{L\overline{X^l_{^\cdot}}\}$, where $[L\overline{X^l_{^\cdot}}]\in\Zz^d$ and each component of  $\{L\overline{X^l_{^\cdot}}\}$ is bounded in modulus by $1/2$, $L\overline{X^l_{^\cdot}}$ can be replaced with $\{L\overline{X^l_{^\cdot}}\}$.
In the notation of [Su12] $q_{n_l}(\overline{X^l_{^\cdot}})=L^d f_{n_l}(0;L\overline{X^l_{^\cdot}})$. Its asymptotic form, valid in the $L\to\infty$ limit was obtained there:
\be\label{qn-asymp}
q_{n_l}(\overline{X^l_{^\cdot}})=
\left\{\begin{array}{lll}
\frac{L^d}{n_l^{d/2}\lambda_{\beta}^{d}}\,[1+o(1)]&\mbox{if}& n_l\lambda_\beta^2/L^2\to 0\\
e^{-\pi n_l\lambda_\beta^2\{L\overline{X^l_{^\cdot}}\}^2/L^2}\left[1
+o(1)\right]&\mbox{if}&n_l\lambda_\beta^2/L^2\to\infty\\
\sum_{z\in\Zz^d}e^{-\pi c(z+\{L\overline{X^l_{^\cdot}}\})^2}&\mbox{if}&n_l\lambda_\beta^2/L^2=c.
\end{array}\right.
\ee
We stress that  on average $q_{n_l}(\overline{X^l_{^\cdot}})$ does not tend to zero when $n_l\lambda_\beta^2/L^2\to\infty$: the analysis in [Su12] showed that $|\overline{X^l_{^\cdot}}|=O(1/\sqrt{n_l})$ for typical sets of $x^k_{j,r}$ when $n_l\to\infty$. Therefore, for $L$ large enough $\{L\overline{X^l_{^\cdot}}\}^2/L^2=\overline{X^l_{^\cdot}}^2=O(1/n_l)$ and
 $e^{-\pi n_l\lambda_\beta^2\overline{X^l_{^\cdot}}^2}$ tends to some positive number.

Finally,
$\delta(X^0_1,\dots,X^p_1)$ restricts the integrals to a submanifold on which each $X^l_1$ is zero, and $K_{\{\alpha^k_j\}}$ is the number of linearly independent equations $X^l_1=0$. To explain $\Delta_{\{\alpha^k_j\},\{n_l\}_0^p}$, we refer to [Su11]: if the $p+1$ cycles are represented by the labelled vertices of a multigraph ${\cal G}_{\{\alpha^k_j\}}$ with $\sum_{j\in C_{l'},\,k\in C_l}\alpha^k_j$ edges between the vertices $l'<l$, then the non-isolated vertices must form mergers of circles of any ($\geq 2$) length. Now
$\Delta_{\{\alpha^k_j\},\{n_l\}_0^p}=1$ if ${\cal G}_{\{\alpha^k_j\}}$ is a merger graph in the above sense, and is zero otherwise.

We first prove the occurrence of cycles whose length increases proportionally to $N$ in a simplified model, retaining from the multiple sum that constitutes $\Phi^N_{n,\{n_l\}_1^p}$ the single term $\alpha^k_j=0$ if $j$ and $k$ are in different cycles. We shall refer to it as the cycle-decoupling model and denote its partition function by $Q_{N,L}^{\rm dcp}$.
For a while we still continue with a general integrable $u$.
Compared to $Q_{N,L}$,
\be
Q_{N,L}^{\rm dcp}=\frac{\exp\left\{-\frac{\beta\hat{u}(0)N(N-1)}{2L^d}\right\}}{N}
\left(\Phi^N_N+\sum_{n=1}^{N-1}\Phi^n_n
\sum_{p=1}^{N-n}\frac{1}{p!}\sum_{\{n_l\}_{l=1}^p\vdash N-n}\prod_{l=1}^p\frac{1}{n_l}\Phi^{n_l}_{n_l}\right).
\ee
Here $\Phi^{n_l}_{n_l}$ is the precise analogue of $\Phi^N_N$, cf.  Eq.~(\ref{PhiNN}), all the particles (to be considered) are in a single cycle.
That is,
\be\label{Phi^{n_l}_{n_l}}
\Phi^{n_l}_{n_l}
=\sum_{\{\alpha^k_j \in\Nn_0| \{j<k\}\subset C_l\}}
\,\prod_{\{j<k\}\subset C_l}
\frac{ \left(-\beta\right)^{\alpha^k_j}}{\alpha^k_j !}
\prod_{r=1}^{\alpha^k_j}\int_0^1\d t^k_{j,r} \int \d x^k_{j,r}\ \hat{u}\left(x^k_{j,r}\right)
e^{-\pi n_l \lambda_\beta^2\left[\overline{\left(X^l_{^\cdot}\right)^2|_l}-\left(\overline{X^l_{^\cdot}|_l}\right)^2\right]}
q_{n_l}(\overline{X^l_{^\cdot}|_l}).
\ee
Note that the only $L$-dependence of $\Phi^{n_l}_{n_l}$ is in $q_{n_l}(\overline{X^l_{^\cdot}|_l})$.
By comparison with the case of $n_l=N$, for L so large that $L^{-d}\sum_{z^k_{j,r}\in\Zz^d\setminus\{0\}}\hat{u}(z^k_{j,r}/L)$ can be replaced with $\int \d x^k_{j,r}\ \hat{u}\left(x^k_{j,r}\right)$,
\bea
\lefteqn{
\int_\Lambda \d x_l \int W_{x_l x_l}^{n_l\beta}(\d\omega_l) e^{-\beta U(\omega_l)}
=
e^{-\beta\hat{u}(0)n_l(n_l-1)/2L^d}     }
\nonumber\\
&&\times
\sum_{\{\alpha^k_j \in\Nn_0| \{j<k\}\subset C_l\}}
\,\prod_{\{j<k\}\subset C_l}
\frac{ \left(-\beta\right)^{\alpha^k_j}}{\alpha^k_j !}
\prod_{r=1}^{\alpha^k_j}\int_0^1\d t^k_{j,r} \int \d x^k_{j,r}\ \hat{u}\left(x^k_{j,r}\right)
e^{-\pi n_l \lambda_\beta^2\left[\overline{\left(X^l_{^\cdot}\right)^2|_l}-\left(\overline{X^l_{^\cdot}|_l}\right)^2\right]}
q_{n_l}(\overline{X^l_{^\cdot}|_l}),
\nonumber\\
\eea
which yields the Feynman-Kac form of $\Phi^{n_l}_{n_l}$,
\be\label{Phi^{n_l}_{n_l}-FK}
\Phi^{n_l}_{n_l}
=e^{\beta\hat{u}(0)n_l(n_l-1)/2L^d}\int_\Lambda \d x_l \int W_{x_l x_l}^{n_l\beta}(\d\omega_l) e^{-\beta U(\omega_l)}.
\ee
We can incorporate the mean-field factor into the partition function by defining
\[
\tilde{Q}_{N,L}^{\rm dcp}:=e^{\frac{\beta\hat{u}(0) N(N-1)}{2L^d}}Q_{N,L}^{\rm dcp}.
\]
Then
\be\label{Qdcp-tilde}
\tilde{Q}_{N,L}^{\rm dcp}
=\frac{1}{N}
\left(\Phi^N_N+\sum_{n=1}^{N-1}\Phi^n_n
\sum_{p=1}^{N-n}\frac{1}{p!}\sum_{\{n_l\}_{l=1}^p\vdash N-n}\prod_{l=1}^p\frac{1}{n_l}\Phi^{n_l}_{n_l}\right)
\ee
and in the average over the partitions of $N-n$ one can recognize $\tilde{Q}_{N-n,L}^{\rm dcp}$. Defining $\tilde{Q}_{0,L}^{\rm dcp}=1$, we have therefore
\be\label{Q-decoupling}
\tilde{Q}_{N,L}^{\rm dcp}=\frac{1}{N}\sum_{n=1}^{N}\Phi^n_{n}\tilde{Q}_{N-n,L}^{\rm dcp}.
\ee
The estimations done in the Lemma for $\int_\Lambda \d x \int W_{x x}^{n\beta}(\d\omega) e^{-\beta U(\omega)}$ provide us with bounds on $\Phi^{n}_{n}$.
First, the superstability bound
\[
U(\omega)
\geq - Bn+C_\Lambda[u] n(n-1)/L^d
\]
implies
\bea\label{upperbound-on-phi^l_nl}
\Phi^n_{n}
\leq q_n  e^{\beta Bn}\,
e^{\beta\left[\hat{u}(0)/2-C_\Lambda[u]\right]n(n-1)/L^d}.
\eea
From Eqs.~(\ref{Jensen}) and (\ref{av-U-final-upper-bound}) we can infer the lower bound
\bea\label{lowerbound-on-phi^l_nl}
\Phi^n_{n}
\geq  q_n e^{-2^{d/2-1}\zeta(d/2)\beta \|u\|_1 n/\lambda_\beta^d}\,
e^{-\beta\left[\|u\|_1-\hat{u}(0)\right] n(n-1)/2L^d}.
\eea
It is at this point that we abandon the general potential and continue with $u\geq 0$ and $\hat{u}\geq 0$, when $\hat{u}(0)=\|u\|_1=2C_\Lambda[u]$, $B=u_L/2$. The upper and lower bounds for $\Phi^n_{n}$ become
\be\label{upperlower-on-phi^l_nl}
q_n e^{-2^{d/2-1}\zeta(d/2)\beta \hat{u}(0) n/\lambda_\beta^d}
\leq\Phi^n_{n}
\leq
q_n  e^{\beta u_L(0)n/2}.
\ee
Upper and lower bounds for $\tilde{Q}_{N,L}^{\rm dcp}$ can be obtained by substituting (\ref{upperlower-on-phi^l_nl}) into Eq.~(\ref{Qdcp-tilde}):
\be\label{Qdcp-tilde-upper}
e^{-2^{d/2-1}\zeta(d/2)\beta \hat{u}(0) N/\lambda_\beta^d}Q^0_{N,L}
\leq
\tilde{Q}_{N,L}^{\rm dcp}
\leq e^{\beta u_L(0)N/2} Q^0_{N,L}\qquad (u\geq 0,\ \hat{u}\geq 0).
\ee
These are the same as those for $\tilde{Q}_{N,L}=e^{\frac{\beta\hat{u}(0) N(N-1)}{2L^d}}Q_{N,L}$, so any modification of $\tilde{Q}_{N,L}^{\rm dcp}$ due to coupling must fit within these bounds.
Equation~(\ref{Q-decoupling}) has the same form as (\ref{Q0NL}) for the partition function of the ideal Bose gas and, given $\Phi^n_{n}$, it defines $\tilde{Q}_{N,L}^{\rm dcp}$ recursively.
We can proceed as we did above for the ideal gas.
Writing $\tilde{Q}_{N,L}^{\rm dcp}=\exp\{-\beta L^d \tilde{f}_{N,L}^{\rm dcp}\}$, if $N$ and $L$ are large and $n=o(N)$ then
\be
\frac{\tilde{Q}_{N-n,L}^{\rm dcp}}{\tilde{Q}_{N,L}^{\rm dcp}}=\exp\left\{\beta n\frac{\tilde{f}_{N,L}^{\rm dcp}-\tilde{f}_{N-n,L}^{\rm dcp}}{(n/L^d)}\right\}
= \exp\left\{\beta n[\partial \tilde{f}^{\rm dcp}/\partial \rho+o(1)]\right\}.
\ee
Here we anticipate that $\partial \tilde{f}^{\rm dcp}/\partial \rho$ exists. It is the chemical potential of the cycle-decoupling model, and below it will be obtained as the solution of Eq.~(\ref{mu=tilde-f}).
Let $g_N=o(N^{2/d})$ be a sequence of positive integers that tends to infinity. Repeating the argument given in the discussion of the ideal gas,
\bea\label{lim1-dcp}
\lim_{N,L\to\infty, N/L^d=\rho}\frac{1}{N}\sum_{n=1}^{g_N}\Phi^n_{n} \frac{\tilde{Q}_{N-n,L}^{\rm dcp}}{\tilde{Q}_{N,L}^{\rm dcp}}
=\frac{1}{\rho\lambda_\beta^d}\sum_{n=1}^\infty \frac{\varphi^0_{n}\exp\left\{\beta n\partial \tilde{f}^{\rm dcp}/\partial \rho\right\}}{n^{d/2}}\leq 1
\eea
where $\varphi^0_1=1$ and for $n\geq 2$
\bea
\varphi^0_{n}
&=&
\sum_{\{\alpha^k_j \in\Nn_0| 1\leq j<k\leq n\}}
\frac{ \left(-\beta\right)^{\alpha^k_j}}{\alpha^k_j !}
\prod_{r=1}^{\alpha^k_j}\int_0^1\d t^k_{j,r} \int \d x^k_{j,r}\ \hat{u}\left(x^k_{j,r}\right)
e^{-\pi n \lambda_\beta^2\left[\overline{\left(X^0_{^\cdot}\right)^2|_0}-\left(\overline{X^0_{^\cdot}|_0}\right)^2\right]}.
\eea
In Eq.~(\ref{lim1-dcp}) it was possible to replace $\Phi^n_n/N$ with $\varphi^0_{n}/(\rho\lambda_\beta^d n^{d/2})$ because $n\lambda_\beta^2/L^2\to 0$ and thus
$q_{n}(\overline{X^0_{^\cdot}|_0})\sim L^d/(\lambda_\beta^d n^{d/2})$ as $L$ increases, cf. Eq.~(\ref{qn-asymp}).
As in Eq.~(\ref{lim1}), the limit of the sum up to $g_N$ contains all the contribution of finite cycles and nothing else, the sum from $M$ to $g_N$ goes to zero if first $N$ and then $M$ tends to infinity; the consequence is that the asymptotic weight of cycles whose length diverges with $N$ but the divergence is slower than $N^{2/d}$ is zero.
The bounds (\ref{upperlower-on-phi^l_nl}) show that $\varphi^0_{n}$ cannot increase or decrease faster than exponentially in $n$.
Thus, $e^{-b_ln}\leq\varphi^0_n\leq e^{b_un}$ with positive numbers $b_l$, $b_u$.
Because $\varphi^0_{n}$ is independent of the density, the infinite sum in (\ref{lim1-dcp}) can increase with $\rho$ only due to an increase of $\partial \tilde{f}^{\rm dcp}/\partial \rho$. However, $-\limsup_{n\to\infty}n^{-1}\ln\varphi^0_n\leq b_l$ is an upper bound to
$\beta\partial \tilde{f}^{\rm dcp}/\partial \rho$, otherwise the infinite sum could not be convergent for $\rho$ large enough. It follows that
\be\label{sup}
\sup_\rho\sum_{n=1}^\infty \frac{\varphi^0_{n}\exp\left\{\beta n\partial \tilde{f}^{\rm dcp}/\partial \rho\right\}}{n^{d/2}}
=\zeta^{\rm dcp}(\beta)<\infty.
\ee
If $\rho\leq\zeta^{\rm dcp}(\beta)/\lambda_\beta^d$, the equation
\be\label{mu=tilde-f}
\frac{1}{\rho\lambda_\beta^d}\sum_{n=1}^\infty \frac{\varphi^0_{n}\exp\{\beta n\mu\}}{n^{d/2}}=1
\ee
has a unique solution for $\mu$ that can be identified with $\partial \tilde{f}^{\rm dcp}/\partial \rho$. It is an increasing function of $\rho$ which reaches its maximum $\bar{\mu}=\bar{\mu}(\beta)$ at $\rho=\zeta^{\rm dcp}(\beta)/\lambda_\beta^d$, where $\bar{\mu}$ satisfies the equation
\be
\frac{1}{\zeta^{\rm dcp}(\beta)}\sum_{n=1}^\infty \frac{\varphi^0_{n}\exp\{\beta n\bar{\mu}\}}{n^{d/2}}=1.
\ee
So $\partial \tilde{f}^{\rm dcp}(\rho,\beta)/\partial \rho\equiv\bar{\mu}(\beta)$ if $\rho\geq\zeta^{\rm dcp}(\beta)/\lambda_\beta^d$.
From [Su12],
\[
\lim_{n\to\infty}n^{-1}\ln\varphi^0_n= c  e^{-\epsilon\beta}
\]
with some positive constants $c$ and $\epsilon$. Hence, $\bar{\mu}(\beta)=-c\, e^{-\epsilon\beta}/\beta$.
If $\varphi^0_n$ were a pure exponential then $\zeta^{\rm dcp}(\beta)=\zeta(d/2)$ would be, with the consequence that the critical density would agree with that of the ideal Bose gas. However, $\varphi^0_n$ is not a pure exponential.
If $\rho>\zeta^{\rm dcp}(\beta)/\lambda_\beta^d$, in Eq.~(\ref{lim1-dcp}) the inequality is strict, and the completion to 1 comes from cycle lengths diverging at least as fast as $N^{2/d}$.
More is true, however. Choose any $\varepsilon>0$, then
\be
\lim_{N\to\infty}\frac{1}{N}\sum_{n=\lfloor\varepsilon N^{2/d}\rfloor}^N 1=1,\quad
\lim_{N,L\to\infty, N/L^d=\rho}\frac{1}{N}\sum_{n=\lfloor\varepsilon N^{2/d}\rfloor}^N \Phi^n_{n} \frac{\tilde{Q}_{N-n,L}^{\rm dcp}}{\tilde{Q}_{N,L}^{\rm dcp}}<1.
\ee
This tells us that on average $\Phi^n_{n}\,\tilde{Q}_{N-n,L}^{\rm dcp}/\tilde{Q}_{N,L}^{\rm dcp}<1$ if $n\geq\varepsilon N^{2/d}$, which suggests that
for any $h_N=o(N)$, $h_N>\varepsilon N^{2/d}$
\be\label{no-submacroscopic}
\lim_{N,L\to\infty, N/L^d=\rho}\frac{1}{N}\sum_{n=\lfloor\varepsilon N^{2/d}\rfloor}^{h_N} \Phi^n_{n} \frac{\tilde{Q}_{N-n,L}^{\rm dcp}}{\tilde{Q}_{N,L}^{\rm dcp}}=0.
\ee
Now $q_{n}(\overline{X^0_{^\cdot}|_0})=O(1)$ for $n\geq \varepsilon N^{2/d}$, so $\Phi^n_{n}\leq C\varphi^0_n$ with some constant $C$, and $\varphi^0_n$ is independent of $N$. At the same time
\[
\tilde{Q}_{N-n,L}^{\rm dcp}/\tilde{Q}_{N,L}^{\rm dcp}\sim e^{n\beta\bar{\mu}(\beta)},
\]
also independent of $N$. Thus, we are dealing with a sequence $a_n>0$ such that
\bea\label{a_n}
\lim_{N\to\infty}\frac{1}{N}\sum_{n=\lfloor\varepsilon N^{2/d}\rfloor}^N a_n
&=&\lim_{N\to\infty}\frac{N-\varepsilon N^{2/d}}{N}\frac{1}{N-\varepsilon N^{2/d}}\sum_{n=\lfloor\varepsilon N^{2/d}\rfloor}^N a_n
\nonumber\\
&=&\lim_{N\to\infty}\frac{1}{N-\varepsilon N^{2/d}}\sum_{n=\lfloor\varepsilon N^{2/d}\rfloor}^N a_n \leq 1.
\eea
Supposing the opposite of (\ref{no-submacroscopic}),
\[
\lim_{N\to\infty}\frac{1}{N}\sum_{n=\lfloor\varepsilon N^{2/d}\rfloor}^{h_N}a_n
= \lim_{N\to\infty}\frac{h_N-\varepsilon N^{2/d}}{N}\frac{1}{h_N-\varepsilon N^{2/d}}
\sum_{n=\lfloor\varepsilon N^{2/d}\rfloor}^{h_N}a_n>0.
\]
This implies
\[
\lim_{N\to\infty}\frac{1}{h_N-\varepsilon N^{2/d}}
\sum_{n=\lfloor\varepsilon N^{2/d}\rfloor}^{h_N}a_n
=\lim_{M\to\infty}\frac{1}{M-m}\sum_{n=m}^{M}a_n=\infty,
\]
in contradiction with (\ref{a_n}).
Therefore (\ref{no-submacroscopic}) holds true, and
\be\label{part-1-decoupled}
\frac{\zeta^{\rm dcp}(\beta)}{\rho\lambda_\beta^d}
+
\lim_{\varepsilon\downarrow 0}\lim_{N,L\to\infty, N/L^d=\rho}\frac{1}{N}
\sum_{n=\lfloor \varepsilon N\rfloor}^N\Phi^n_n \frac{\tilde{Q}_{N-n,L}^{\rm dcp}}{\tilde{Q}_{N,L}^{\rm dcp}}=1   \qquad (\rho>\zeta^{\rm dcp}(\beta)/\lambda_\beta^d).
\ee
As in the noninteracting gas, in infinite volume the cycles are either finite or macroscopic.
Note that apart from Eq.~(\ref{qn-asymp}) we did not use the Fourier-expanded form of the partition function.

\vspace{5pt}
Now consider the full model. We write the partition functions in their form not distinguishing cycle 0,
\be
\tilde{Q}_{N,L}=\sum_{p=0}^N\frac{1}{p!}\sum_{\{n_l\}_0^p\vdash N}\frac{1}{\prod_{l=0}^p n_l}\Phi^N_{\{n_l\}_0^p},
\qquad
\tilde{Q}_{N,L}^{\rm dcp}=\sum_{p=0}^N\frac{1}{p!}\sum_{\{n_l\}_0^p\vdash N}\frac{1}{\prod_{l=0}^p n_l}\prod_{l=0}^p \Phi^{n_l}_{n_l}.
\ee
Let a partition $\{n_l\}_{l=0}^p\vdash N$ be given.
$\Delta_{\{\alpha^k_j\},\{n_l\}_0^p}$ and $K_{\{\alpha^k_j\}}$ depend only on
\[
~\alpha^{\rm cp}=\{\alpha^k_j\in\Nn_0| \mbox{$j$ and $k$ are in different cycles}\}.
\]
For the set of $~\alpha^{\rm cp}$ that satisfy $\Delta_{\{\alpha^k_j\},\{n_l\}_0^p}=1$ we use the notation
$
A^{\rm cp}_{\{n_l\}_0^p},
$
and the $d\left(\sum_{\alpha^k_j\in~\alpha^{\rm cp}}\alpha^k_j-K_{\{\alpha^k_j\}}\right)$ dimensional manifold in $\Rr^{d\sum_{\alpha^k_j\in~\alpha^{\rm cp}}\alpha^k_j}$ on which $X^0_1=X^1_1=\cdots=X^p_1=0$ will be denoted by
$X_{~\alpha^{\rm cp}}$. There is no constraint for
\[
~\alpha_l=\{\alpha^k_j\in\Nn_0| \{j<k\}\subset C_l\}.
\]
Also, let
\be
\|~\alpha_l\|=\sum_{\alpha^k_j\in~\alpha_l}\alpha^k_j,\qquad
\|~\alpha^{\rm cp}\|=\sum_{\alpha^k_j\in~\alpha^{\rm cp}}\alpha^k_j,\qquad
\|\hat{u}^{~\alpha^{\rm cp}}\|= \int_{X_{~\alpha^{\rm cp}}}\prod_{j,k,r}\hat{u}(x^k_{j,r})\prod_{j,k,r}\d x^k_{j,r}.
\ee
The coupling brings in $\|~\alpha^{\rm cp}\|$ factors $\beta$ and $\hat{u}$, and $\beta^{\|~\alpha^{\rm cp}\|}\,\|\hat{u}^{~\alpha^{\rm cp}}\|$ is a volume raised to the power $K_{~\alpha^{\rm cp}}$. Denoting this volume by $v_{\beta,~\alpha^{\rm cp}}$,
\be
\beta^{\|~\alpha^{\rm cp}\|}\,\|\hat{u}^{~\alpha^{\rm cp}}\|=(v_{\beta,~\alpha^{\rm cp}})^{K_{~\alpha^{\rm cp}}}.
\ee
With these notations
\bea\label{Phi^N_n-new-notation}
\Phi^N_{\{n_l\}_0^p}
=
\prod_{l=0}^p \sum_{~\alpha_l}
\frac{[-\beta u(0)]^{\|~\alpha_l\|}}{\prod_{\alpha^k_j\in~\alpha_l}\alpha^k_j !}
\prod_{\alpha^k_j\in~\alpha_l}
\prod_{r=1}^{\alpha^k_j}\int_0^1\d t^k_{j,r}
\int_{\Rr^{d}}\d x^k_{j,r}\frac{\hat{u}(x^k_{j,r})}{u(0)}
e^{-\pi n_l \lambda_\beta^2\left[\overline{\left(X^l_{^\cdot}\right)^2|_l}-\left(\overline{X^l_{^\cdot}|_l}\right)^2\right]}
q_{n_l}(\overline{X^l_{^\cdot}|_l})
\nonumber\\
\sum_{~\alpha^{\rm cp}\in A^{\rm cp}_{\{n_l\}_0^p}}
\frac{(-1)^{\|~\alpha^{\rm cp}\|}}{\prod_{\alpha^k_j\in ~\alpha^{\rm cp}}\alpha^k_j !}
\left[\frac{\rho v_{\beta,~\alpha^{\rm cp}}}{N}\right]^{K_{~\alpha^{\rm cp}}}
\int_{X_{~\alpha^{\rm cp}}}\frac{\prod_{j,k,r}\d x^k_{j,r}\hat{u}(x^k_{j,r})}{\|\hat{u}^{~\alpha^{\rm cp}}\|}
\prod_{\alpha^k_j\in ~\alpha^{\rm cp}}\prod_{r=1}^{\alpha^k_j}\int_0^1\d t^k_{j,r}
\nonumber\\
\exp\left\{-\pi \lambda_\beta^2\sum_{l=0}^p n_l\left[
\overline{\left(X^l_{^\cdot}\right)^2}-\overline{X^l_{^\cdot}}^2-\left(\overline{\left(X^l_{^\cdot}|_{l}\right)^2}-\left(\overline{X^l_{^\cdot}|_{l}}\right)^2  \right) \right]\right\}
\prod_{l=0}^p\frac{q_{n_l}(\overline{X^l_{^\cdot}})}{q_{n_l}(\overline{X^l_{^\cdot}|_l })}.
\eea
Observe that
\be\label{averages}
\int_{X_{~\alpha^{\rm cp}}}\prod_{j,k,r}\d x^k_{j,r} \frac{\prod_{j,k,r}\hat{u}(x^k_{j,r})}{\|\hat{u}^{~\alpha^{\rm cp}}\|}
\prod_{\alpha^k_j\in ~\alpha^{\rm cp}}\prod_{r=1}^{\alpha^k_j}\int_0^1\d t^k_{j,r} =1,\qquad
\prod_{\alpha^k_j\in~\alpha_l}
\prod_{r=1}^{\alpha^k_j}\int_0^1\d t^k_{j,r}\int_{\Rr^d}\d x^k_{j,r}\frac{\hat{u}(x^k_{j,r})}{u(0)}=1
\ee
($u(0)=\|\hat{u}\|_1$.) In Eq.~(\ref{Phi^N_n-new-notation}) the first line alone is $\prod_{l=0}^p\Phi^{n_l}_{n_l}$.
The vectors $x^k_{j,r}$ for $j$ and $k$ in the same cycle are independent identically distributed random variables; those with $j$ and $k$ in different cycles follow a coupled distribution. The probability measures can be read off from (\ref{averages}).
The sum over $A^{\rm cp}_{\{n_l\}_0^p}$ and the coupled distribution still factorize according to the maximal connected components of cycles.
The sign of a term depends only on the parity of $\|~\alpha^{\rm cp}\|$ and, because $\Phi^N_{\{n_l\}_0^p}>0$, the positive terms dominate.
Otherwise, the inclusion of $~\alpha^{\rm cp}$ has a threefold effect:
(i) the fluctuations of the momenta $\overline{\left(X^l_{^\cdot}\right)^2}-\overline{X^l_{^\cdot}}^2$ increase on average, (ii) For $K_{~\alpha^{\rm cp}}>0$ there appears a dimensionless factor $(\rho v_{\beta,~\alpha^{\rm cp}}/N)^{K_{~\alpha^{\rm cp}}}$, and (iii) the large number of different ways to couple the cycles contributes to the entropy.
The analysis in [Su12] showed that in the infinite-volume limit of ${\cal G}_{~\alpha^{\rm cp}}$ each vertex is of finite degree, i.e. the exponentially dominant contribution to the partition function comes from terms in which $\sum_{j\in C_l}\left[\sum_{l'>l}\sum_{k\in C_{l'}}\alpha^k_j+\sum_{l'<l}\sum_{k\in C_{l'}}\alpha^j_k\right]$ remains finite for every $l$ as $N$ tends to infinity.
A trivial reason is the rapid decrease of $1/\alpha^k_j!$ for any given pair $(j, k)$; a nontrivial reason is that
keeping the number of edges incident on every vertex finite the increase of fluctuations can be controlled. For $~\alpha^{\rm cp}$ thus chosen the entropy wins the competition between (ii) and (iii) for all $p$ of order $N$ and all values of $\rho$ and $\beta$, but the increase of $\overline{\left(X^l_{^\cdot}\right)^2}-\overline{X^l_{^\cdot}}^2$ influences the dependence on them. In a first time we disregard this latter and focus on the interplay between (ii) and (iii).

Let us recall from [Su11] that a connected merger graph of $V$ vertices contributes $V-1$ to $K_{~\alpha^{\rm cp}}$. So if the coupling regroups the $p+1$ vertices (each representing a cycle) into $m_{~\alpha^{\rm cp}}$ connected components then $K_{~\alpha^{\rm cp}}=p+1-m_{~\alpha^{\rm cp}}$. The connected components are isolated vertices and clusters (circles composed of two or more vertices and their mergers).
Let $N_{\rm isl}$ be the number of isolated vertices, $N_{\rm cls}$ the number of clusters and $N_{\rm nisl}$ the total number of non-isolated vertices (i.e. those in clusters).
Then $p+1=N_{\rm isl}+N_{\rm nisl}$, $m_{~\alpha^{\rm cp}}=N_{\rm isl}+N_{\rm cls}$ and
\[
K_{~\alpha^{\rm cp}}=N_{\rm nisl}-N_{\rm cls}=\sum_{i=1}^{N_{\rm cls}}(V_i-1)
\]
where $V_i\geq 2$ is the number of vertices in the $i$th cluster. We must analyze partial sums of $A^{\rm cp}_{\{n_l\}_0^p}$ running over $~\alpha^{\rm cp}$ such that $\|~\alpha^{\rm cp}\|$ is even and increases linearly with $N_{\rm nisl}$.
Our aim is to show that the coupling among the cycles produces a global factor $e^{CN}$ ($C>0$) that appears in $\tilde{Q}_{N,L}$ compared to $\tilde{Q}_{N,L}^{\rm dcp}$.
We demonstrate the presence of such a factor on an example taken over from [Su11].

Let $p+1=cN$ where $c<1$.
Let $N_{\rm isl}=aN$ ($0<a<c$) and suppose that the remaining $N_{\rm nisl}=(c-a)N$ vertices are coupled in $N_{\rm cls}=\frac{1}{2}N_{\rm nisl}$ two-circles. Then $m_{~\alpha^{\rm cp}}=(a+(c-a)/2)N=(c+a)N/2$ and $K_{~\alpha^{\rm cp}}=(c-a)N/2$.
For $\prod_{\alpha^k_j\in ~\alpha^{\rm cp}}(1/\alpha^k_j !)$ we can substitute $\epsilon_0^{N_{\rm nisl}}$ with some $\epsilon_0<1$.
The number of $~\alpha^{\rm cp}$ that differ only in the permutations of the (labelled) vertices is
\[
{N_{\rm isl}+N_{\rm nisl}\choose N_{\rm isl}}\frac{N_{\rm nisl}!}{N_{\rm cls}!\, 2^{N_{\rm cls}}}
=
{cN\choose aN} \frac{[(c-a)N]!}{\left[\frac{(c-a)N}{2}\right]!\,2^{\frac{(c-a)N}{2}}}
\]
so altogether we get
\be\label{loss-times-gain}
\epsilon_0^{N_{\rm nisl}}\left[\frac{\rho v_{\beta,~\alpha^{\rm cp}}}{N}\right]^{N_{\rm nisl}-N_{\rm cls}}\frac{(N_{\rm isl}+N_{\rm nisl})!}{N_{\rm isl}!\, N_{\rm cls}!\,2^{N_{\rm cls}} }
=
\left[\frac{\epsilon\rho v_{\beta,~\alpha^{\rm cp}}}{N}\right]^{\frac{(c-a)N}{2}}\frac{(cN)!}{(aN)!\,\left[\frac{(c-a)N}{2}\right]!\,2^{\frac{(c-a)N}{2}}}
\sim
 \left[\left(\frac{\epsilon\rho v_{\beta,~\alpha^{\rm cp}}}{e(c-a)}\right)^{\frac{c-a}{2}}\frac{c^c}{a^a}\right]^N.
\ee
Here $\epsilon=\epsilon_0^2$.
For any nonzero $\rho v_{\beta,~\alpha^{\rm cp}}$, choosing $a$ so that $0<c-a<\epsilon\rho v_{\beta,~\alpha^{\rm cp}}/e$, the expression above increases exponentially with $N$.
Note also that it equals 1 if $a=c$ (no coupling) and tends to $[c\epsilon\rho v_{\beta,~\alpha^{\rm cp}}/e]^{cN/2}$ if $a$ goes to zero (full coupling). This shows that in order to obtain an exponentially large contribution for arbitrarily small $\rho v_{\beta,~\alpha^{\rm cp}}$ a positive but not full density of uncoupled cycles, i.e. $0<a<c$ is necessary.
Our choice of $(c-a)N/2$ two-circles exemplifies the general case when both $N_{\rm isl}$ and $N_{\rm cls}$ are proportional to $N$.

We still must count with the increase of $\overline{\left(X^l_{^\cdot}\right)^2}-\overline{X^l_{^\cdot}}^2$ appearing in Eq.~(\ref{Phi^N_n-new-notation}) in the mean value of
\bea\label{origin-corr-factor}
\exp\left\{
-\pi\lambda_\beta^2\sum_{l=0}^p n_l\left[
\overline{\left(X^l_{^\cdot}\right)^2}-\overline{X^l_{^\cdot}}^2-\left(\overline{\left(X^l_{^\cdot}\right)^2|_{l}}-\left(\overline{X^l_{^\cdot}|_{l}}\right)^2  \right) \right]\right\}
\prod_{l=0}^p\frac{q_{n_l}(\overline{X^l_{^\cdot}})}{q_{n_l}(\overline{X^l_{^\cdot}|_l })}
\hspace{5.5cm}
\nonumber\\
=
\exp\left\{
\sum_{l=0}^p\left[-\pi\lambda_\beta^2 n_l
\left(\overline{\left(X^l_{^\cdot}\right)^2|_{\neg l}} - \left(\overline{X^l_{^\cdot}|_{\neg l}}\right)^2 +2\,\left[\overline{X^l_{^\cdot}|_l \cdot X^l_{^\cdot}|_{\neg l}}- \overline{X^l_{^\cdot}|_l} \cdot \overline{X^l_{^\cdot}|_{\neg l}}\right]\right)
+\ln q_{n_l}(\overline{X^l_{^\cdot}})-\ln q_{n_l}(\overline{X^l_{^\cdot}|_l }) \right] \right\}.
\nonumber\\
\eea
The mean value of the above expression can be represented by
\be\label{corr-factor}
e^{-c_1 N_{\rm nisl}\lambda_\beta^2\rho^{2/d}}
=\left[ e^{-c_1\lambda_\beta^2\rho^{2/d}} \right]^{(c-a)N}
\ee
with a suitably chosen positive constant $c_1$.
This is explained as follows.\\
(i) In the exponent $n_l$ multiplies averages that involve division with $n_l$, see Eq.~(\ref{Xav-decomposition}), so it only neutralizes this division in $\overline{\left(X^l_{^\cdot}\right)^2}$, $\overline{\left(X^l_{^\cdot}\right)^2|_{l}}$, $\overline{\left(X^l_{^\cdot}\right)^2|_{\neg l}}$ and $\overline{X^l_{^\cdot}|_l \cdot X^l_{^\cdot}|_{\neg l}}$, and leaves behind a factor $1/n_l$ in the other terms. \\
(ii) $q_{n_l}(\overline{X^l_{^\cdot}})/q_{n_l}(\overline{X^l_{^\cdot}|_l })$ tends to 1 if $n_l$ remains finite or increases slower than $L^2$ as $L\to\infty$. On the other hand, $\ln q_{n_l}(\overline{X^l_{^\cdot}})-\ln q_{n_l}(\overline{X^l_{^\cdot}|_l })\approx -\pi n_l\lambda_\beta^2\left(\overline{X^l_{^\cdot}}^2-\overline{X^l_{^\cdot}|_l }^2\right)$ is of order 1 if $n_l$ increases faster than $L^2$, cf. Eq.~(\ref{qn-asymp}) and the comment thereafter.\\
(iii) Uncoupled cycles give zero, therefore the number of terms contributing to the sum is proportional to $N_{\rm nisl}$.\\
(iv)
The $\rho$-dependence can be found by a physical argument.
In the exponent $\lambda_\beta^2$ must be divided with a squared length, and the only relevant length here is the mean distance between neighboring cycles which scales as $\rho^{-1/d}$.
In $~\alpha^{\rm cp}$ there is no information about the spatial position of the pairs that appear with $\alpha^k_j>0$, but such an information is present in $x^k_{j,r}$, the dual-space vector associated with the difference of the position vectors of particles $j$ and $k$. The only physically meaningful interpretation of $\|~\alpha^{\rm cp}\|\propto N_{\rm nisl}$ is that the clusters are formed by neighboring cycles, and the larger the density, the stronger is the coupling of a cycle to its neighbors.
A deeper reason is that the exponent originates from the kinetic energy, cf. [Su11], the repulsive interaction has the tendency to confine the particles, and due to the uncertainty principle the kinetic energy increases with $\rho$ as $\rho^{2/d}$.

To fully account for the effect of coupling (\ref{corr-factor}) must still multiply (\ref{loss-times-gain}). The result is
\bea\label{corr-1}
\mbox{(\ref{loss-times-gain})$\times$(\ref{corr-factor})}
 &=&
 \left[\left(\frac{\epsilon\rho v_{\beta,~\alpha^{\rm cp}}}{e(c-a)}\right)^{\frac{c-a}{2}}\frac{c^c}{a^a}
 \left[e^{-c_1\lambda_\beta^2\rho^{2/d}} \right]^{c-a}\right]^N.
\eea
This is exponentially increasing if
\[
\sqrt{\frac{\epsilon\rho v_{\beta,~\alpha^{\rm cp}}}{c-a}} \left(\frac{c^c}{a^a} \right)^{\frac{1}{c-a}}e^{-c_1\lambda_\beta^2\rho^{2/d}-\frac{1}{2}} >1,
\]
which can be attained for any $\beta$ and $\rho$ by choosing $a<c$ close enough to $c$. Because $c^c/a^c>1$, it holds for
\[
c-a\leq  \epsilon\rho v_{\beta,~\alpha^{\rm cp}}e^{-2c_1\lambda_\beta^2\rho^{2/d}-1}.
\]
The expression (\ref{corr-1}) has a maximum as a function of $a$ somewhere between 0 and $c$. Neglecting $c^c/a^a$, (\ref{corr-1}) is maximal at
\[
c-a=\epsilon\rho v_{\beta,~\alpha^{\rm cp}}e^{-2(c_1\lambda_\beta^2\rho^{2/d}+1)}.
\]
With the maximizing $c-a$ one obtains the exponential factor by which $\tilde{Q}_{N,L}$ exceeds $\tilde{Q}_{N,L}^{\rm dcp}\,$:
\be\label{QNL-increased-1}
\tilde{Q}_{N,L}
=e^{CN} \tilde{Q}_{N,L}^{\rm dcp},
\qquad\qquad
C=\frac{1}{2}\epsilon \rho v_{\beta,~\alpha^{\rm cp}}e^{-2(c_1\lambda_\beta^2\rho^{2/d}+1)}.
\ee
From the point of view of BEC the precise $\rho$- and $\beta$-dependence of $C$ is unimportant.
Dividing
\be
\tilde{Q}_{N,L}
=
e^{CN}\tilde{Q}_{N,L}^{\rm dcp}=e^{CN}\frac{1}{N}\sum_{n=1}^{N}\Phi^n_{n}\tilde{Q}_{N-n,L}^{\rm dcp}
\ee
with $\tilde{Q}_{N,L}$ we are back to the equation
\be
1=\frac{1}{N}\sum_{n=1}^N\Phi^n_{n} \frac{\tilde{Q}_{N-n,L}^{\rm dcp}}{\tilde{Q}_{N,L}^{\rm dcp}}
\ee
and the result about BEC in the cycle-decoupling model. In particular, $\zeta_c(\beta)=\zeta^{\rm dcp}(\beta)$, the critical density $\zeta_c(\beta)/\lambda_\beta^d$ for the full model is the same as for the cycle-decoupling model. What changes is the free energy and the relation between the density and the chemical potential, just as with the inclusion of the mean-field term. However, contrary to the latter there can be a nonanalytic change during BEC also in the inter-cycle contribution to $f(\rho,\beta)$.
This ends the proof of the theorem.

\newsec{Physical meaning of the permutation cycles}

Here we propose a possible interpretation of the permutation cycles as physical entities.
When in the stochastic description $n$ physical particles form an effective single-particle trajectory then quantum-mechanically they occupy one and the same one-particle state. About the nature of this state we can be guided by the analysis of the noninteracting gas and by its partition function
\be
Q^0_{N,L}=\sum_{p=1}^N
\frac{1}{p!}\sum_{n_1,\dots,n_p\geq 1:\sum_1^p n_l=N}\ \prod_{l=1}^p\frac{1}{n_l}
\sum_{z\in\Zz^d}\exp\left\{-\frac{\pi n_l \lambda_\beta^2}{L^2}z^2\right\}.
\ee
Given $\{n_l\}_1^p$ and uniformly distributed random vectors $\{y_l\}_1^p$ in $\Lambda$, all the $n_l$ particles of the $l$th cycle are in a superposition of the plane wave states $L^{-d/2}e^{\i \frac{2\pi}{L}z\cdot (x_l-y_l)}$ with Gaussian weights $\exp\{-\pi n_l\lambda_\beta^2 z^2/(2L^2)\}$,
\bea
\psi^{L}_{n_l,y_l}(x_l)
=\frac{
\sum_{z\in\Zz^d}\exp\left\{-\frac{\pi n_l\lambda_\beta^2}{2L^2} z^2\right\}L^{-d/2}e^{\i \frac{2\pi}{L}z\cdot (x_l-y_l)}}
{\left[\sum_{z\in\Zz^d}\exp\left\{-\frac{\pi n_l\lambda_\beta^2}{L^2} z^2\right\}  \right]^{1/2}}
=\frac{
\frac{2^{d/2}}{(\sqrt{n_l}\lambda_\beta)^d}
\sum_{z\in\Zz^d}\exp\left\{-\frac{2\pi(x_l-y_l+Lz)^2}{n_l\lambda_\beta^2}\right\}}
{\left[L^{-d}\sum_{z\in\Zz^d}\exp\left\{-\frac{\pi n_l\lambda_\beta^2}{L^2} z^2\right\}\right]^{1/2}}
\nonumber\\
=\frac{
\frac{2^{d/2}}{(\sqrt{n_l}\lambda_\beta)^d}\sum_{z\in\Zz^d}\exp\left\{-\frac{2\pi(x_l-y_l+Lz)^2}{n_l\lambda_\beta^2}\right\}}
{(\sqrt{n_l}\lambda_\beta)^{-d/2}\left[\sum_{z\in\Zz^d}\exp\left\{-\frac{\pi L^2z^2}{n_l\lambda_\beta^2}\right\}\right]^{1/2}}
=
\left(\frac{2}{\sqrt{n_l}\lambda_\beta}\right)^{d/2}
\frac{\sum_{z\in\Zz^d}\exp\left\{-\frac{2\pi(x_l-y_l+Lz)^2}{n_l\lambda_\beta^2}\right\}}
{\left[\sum_{z\in\Zz^d}\exp\left\{-\frac{\pi L^2z^2}{n_l\lambda_\beta^2}\right\}\right]^{1/2}}.
\nonumber\\
\eea
The thermal equilibrium state (density matrix) of the noninteracting Bose gas is a mixed state of the form
\be
{\cal D}^0_{N,L}=\sum_{p=1}^N \frac{1}{p!}\sum_{n_1,\dots,n_p\geq 1:\sum_1^p n_l=N}\ \frac{1}{\prod_{l=1}^pn_l}
\bigotimes_{l=1}^p L^{-d}\int_\Lambda |\psi^{L}_{n_l,y_l}\rangle\langle \psi^{L}_{n_l,y_l}|\, \d y_l
\ee
If $\rho<\zeta(d/2)/\lambda_\beta^d$, the cycle lengths remain finite in the infinite system, and
\be\label{psi^{infty}_{n_l,y_l}(x_l)}
\lim_{L\to\infty}\psi^{L}_{n_l,y_l}(x_l)=\left(\frac{2}{\sqrt{n_l}\lambda_\beta}\right)^{d/2}
\exp\left\{-\frac{2\pi(x_l-y_l)^2}{n_l\lambda_\beta^2}\right\},
\ee
the $l$th cycle represents a Gaussian localized state of width $\propto\lambda_{n_l\beta}$ centered at $y_l$. If $\rho>\zeta(d/2)/\lambda_\beta^d$, some $n_l$ will diverge proportionally to $N$. As seen on its first form, $\psi^{L}_{n_l,y_l}(x_l)$ tends to the zero momentum plane wave,
$
\psi^{L}_{n_l,y_l}(x_l)\sim L^{-d/2}
$
as $L$ increases.

In the interacting Bose gas $\prod_{l=1}^p q_{n_l}$ is replaced with $\Phi^N_{\{n_l\}_1^p}$,
\[
\tilde{Q}_{N,L}
=\sum_{p=1}^N \frac{1}{p!}\sum_{n_1,\dots,n_p\geq 1:\sum_1^p n_l=N} \frac{1}{\prod_{l=1}^p n_l}\
\Phi^N_{\{n_l\}_1^p}.
\]
However, once $\{\alpha^k_j, x^k_{j,r}, t^k_{j,r}\}$ are given, for every $l$ it suffices to consider
\[
\sum_{z\in\Zz^d} \exp\left\{-\frac{\pi \lambda_\beta^2}{L^2}\sum_{q\in C_l}\int_0^1\left[z+Z_q(t)\right]^2\d t\right\}
=
\exp\left\{-\pi n_l \lambda_\beta^2\left[\overline{\left(X^l_{^\cdot}\right)^2}-\overline{X^l_{^\cdot}}^2\right]\right\}
\sum_{z\in\Zz^d}\exp\left\{-\pi n_l \lambda_\beta^2\left(\frac{z}{L}+\overline{X^l_{^\cdot}}\right)^2\right\},
\]
see [Su11, Equations (1.9), (1.10)]. Here $C_l=\{N_{l-1}+1,\dots,N_l\}$, $N_l-N_{l-1}=n_l$ as introduced in Section 2. In [Su11] we already interpreted $\hbar(2\pi/L)Z_q(t)=hX_q(t)$ as the shift due to interactions of the momentum of the $q$th particle at "time" $t$ compared to its value in the ideal gas, and $\overline{X^l_{^\cdot}}$ is the average over time and particles of $X_q(t)$. Accordingly, the common wave function of the $n_l$ particles in cycle $l$ is
\bea
\psi^{L,\overline{X^l_{^\cdot}}}_{n_l,y_l}(x_l)
&=&\frac{\sum_{z\in\Zz^d}\exp\left\{-\frac{\pi n_l\lambda_\beta^2}{2L^2} (z+L\overline{X^l_{^\cdot}})^2\right\}L^{-d/2}e^{\i \frac{2\pi}{L}(z+L\overline{X^l_{^\cdot}})\cdot (x_l-y_l)}}
{\left[\sum_{z\in\Zz^d}\exp\left\{-\frac{\pi n_l\lambda_\beta^2}{L^2} (z+L\overline{X^l_{^\cdot}})^2\right\}  \right]^{1/2}}
\nonumber\\
&=&\left(\frac{2}{\sqrt{n_l}\lambda_\beta}\right)^{d/2}\
\frac{\sum_{z\in\Zz^d}\exp\left\{-\frac{2\pi(x_l-y_l+Lz)^2}{n_l\lambda_\beta^2}\right\}\exp\left\{-\i 2\pi z\cdot L\overline{X^l_{^\cdot}}\right\}}
{\left[\sum_{z\in\Zz^d}\exp\left\{-\frac{\pi L^2z^2}{n_l\lambda_\beta^2}\right\}\cos2\pi z\cdot \left\{L\overline{X^l_{^\cdot}}\right\} \right]^{1/2}}.
\eea
Here we applied the identity
\be
\frac{1}{L^d}\sum_{z\in\Zz^d}e^{-\frac{\pi\lambda^2}{L^2}(z+a)^2}e^{\i\frac{2\pi}{L}z\cdot x}
=\frac{1}{\lambda^d}\sum_{z\in\Zz^d}e^{-\frac{\pi}{\lambda^2}(x+Lz)^2}e^{-\i \frac{2\pi}{L}a\cdot(x+Lz)}.
\ee
With
\be
\Phi^N_{\{n_l\}_1^p}=\sum_{\{\alpha^k_j\in\Nn_0|1\leq j< k\leq N\}}\int_0^1\prod_{j,k,r}\d t^k_{j,r}\int\prod_{j,k,r}\d x^k_{j,r} H\left(\{\alpha^k_j, t^k_{j,r},x^k_{j,r}\}\right),
\ee
where $H$ can be read off from Eq.~(\ref{Phi^N_{n,{n_l}_1^p}}), the density matrix is
\bea
\lefteqn{
{\cal D}_{N,L}
=\sum_{p=1}^N \frac{1}{p!}\sum_{n_1,\dots,n_p\geq 1:\sum_1^p n_l=N}\ \frac{1}{\prod_{l=1}^pn_l}
}\nonumber\\
&&\times\frac{1}{\Phi^N_{\{n_l\}_1^p}} \sum_{\{\alpha^k_j\in\Nn_0|1\leq j< k\leq N\}}\int_0^1\prod_{j,k,r}\d t^k_{j,r}\int\prod_{j,k,r}\d x^k_{j,r} H\left(\{\alpha^k_j, t^k_{j,r},x^k_{j,r}\}\right)
\bigotimes_{l=1}^p L^{-d}\int_\Lambda |\psi^{L,\overline{X^l_{^\cdot}}}_{n_l,y_l}\rangle\langle \psi^{L,\overline{X^l_{^\cdot}}}_{n_l,y_l}|\, \d y_l.
\nonumber\\
\eea
If $\rho<\zeta_c(\beta)/\lambda_\beta^d$, the cycle lengths remain finite in the infinite system, and the wave function in infinite volume reduces to the $z=0$ term (\ref{psi^{infty}_{n_l,y_l}(x_l)}) independent of the interaction. It is only when $n_l\lambda_\beta^2/L^2\to\infty$ and $\overline{X^l_{^\cdot}}=O(1/\sqrt{n_l})$, and thus $L\overline{X^l_{^\cdot}}\to 0$, that the wave function tends to the pure zero-momentum state. These are precisely the conditions that we found for BEC and proved to be met for positive and positive-type pair potentials at densities $\rho> \zeta_c(\beta)/\lambda_\beta^d$.
Although asymptotically the individual cycles represent the same states as in the noninteracting gas, their distribution is different because of their coupling via wave vectors.

\newsec{Historical notes}

In 1924 Bose gave a deceptively simple statistical physical derivation of Planck's radiation formula [Bos]. To obtain the good result all he had to do was to distribute the light quanta among the cells of volume $h^3$ of the phase space somewhat differently than usual. In an endnote the German translator Einstein praised the work and promised to apply its method to an ideal atomic gas, that he indeed did in three papers [E1-3].
One must admire the ingenuity of this step. De Broglie's PhD thesis appeared in the same year and Einstein must have known it; yet, atoms have a mass and an extension, how can many of them occupy the same state?
The surprising result that they can as completely flat waves was not received with much enthusiasm. Distinguished colleagues as Halpern, Schr\"odinger or Smekal had difficulty to understand the new "cell counting" that Bose, and Einstein in [E1], used instead of Boltzmann's. Einstein answered the objections in papers [E2,3] and also in letters, see e.g. [Schr] and the very clear response [E4].
Somewhat later another blow came from Uhlenbeck. To quote London [Lon2],

\noindent
\emph{"This very interesting discovery, however, has not appeared in the textbooks, probably because Uhlenbeck in his thesis {\rm [Uh]} questioned the correctness of Einstein's argument. Since, from the very first, the mechanism appeared to be devoid of any practical significance, all real gases being condensed at the temperature in question, the matter has never been examined in detail; and it has been generally supposed that there is no such condensation phenomenon."}

The regard onto Einstein's work changed in 1938 with the discovery of superfluidity [All, Kap]. Fritz London promptly reacted [Lon1], and in a follow-up paper [Lon2] he detailed his view, that superfluidity must have to do with Bose-Einstein condensation -- the name was coined by him. To support his idea, he computed the critical temperature of the ideal gas with the mass of the He4 atom and the density of liquid helium, and found it not very far off, 1K above the $\lambda$ point separating the He I and He II phases. Prior to that he had to reexamine the controversy between Einstein and Uhlenbeck, and take Einstein's side. In his derivation Einstein arrived at an equation connecting the number of particles $N$ to the chemical potential $\mu$, that one must solve for the latter. $\mu$ appears in an infinite sum over the allowed discrete values of the single-particle momentum. Einstein approximated the sum with an integral and observed the convergence of the integral in three dimensions at $\mu=0$, the largest possible value of the chemical potential: as if $N$ could not go beyond a maximum. At the beginning of his second paper he resolved this paradox by assigning the surplus particles to the zero momentum mode. Uhlenbeck, certainly unaware of Ehrenfest's earlier and identical criticism, argued that without the approximation by an integral the paradox disappears, the original equation can be solved with a $\mu<0$ for arbitrarily large $N$. At that time two crucial mathematical elements were missing from the weaponry of theoretical physicists: a clear notion of the thermodynamic limit and its importance to see a sharp phase transition in the framework of statistical physics, and the appearance of the Dirac delta in probability theory, the fact that in the limit of a sequence of discrete probability distributions an atomic measure can emerge on a continuous background. Einstein's intuition worked correctly, he tacitly performed the thermodynamic limit. The separate treatment of the zero momentum state bothered physicists for a long time, including Feynman, who proposed an alternative derivation based on the statistics of permutation cycles [Fe2]. Simultaneously with London's publications, in a paper written with Kahn, Uhlenbeck also admitted that Einstein was right [Kah]. Tisza published his two-fluid theory about the same time [T1,2], and attributed the specific transport properties of helium II to BEC. However, the argument against describing a strongly interacting dense system of atoms with an ideal gas persisted and set the task: prove BEC in the presence of interaction.

The first consistent theory of superfluidity was given by Landau in 1941 [Lan1]. This extremely influential work denied all connection with BEC (clearly, a position taken against London and even more Tisza, who was earlier in his group in Kharkiv; see also Kadanoff [Kad]). The research on BEC for interacting bosons started after World War II and produced a huge number of papers that we can review only in great lines. Monographs about it and its connection with superfluidity extend over decades, some of them are [Lon4, No, Gr, Sew, Pet, Pi1, Li5, Leg, Ued, Ver, Kag]; those written after 2000 usually cover also the theory of trapped dilute ultra-cold gases of alkaline atoms, that we will not discuss.

Maybe the first and certainly one of the most important contributions was that of Bogoliubov, who described superfluidity on the basis of BEC of weakly interacting bosons [Bog1]; for a recent review see [Z]. His theory had an immense impact on the forthcoming research in quantum statistical physics. Bogoliubov initiated the algebraic approach, the use of second quantization. Writing the Hamiltonian in terms of creation and annihilation operators proved to be very fruitful, because it opened the way for diverse approximations. The interaction, which now must be integrable, appears in a quartic expression. Dropping everything not reducible to a quadratic form makes it possible to diagonalize the Hamiltonian. These models, including those expressible with number operators and known under the names of mean-field, imperfect, perturbed mean-field, full diagonal Bose gas model, lead naturally to BEC [Hu4, Dav, Fa1, Buf, Ber1-Ber4, Lew1, Do1, Do2]. (For a discussion of some problems arising with the truncation of the Hamiltonian see the Introduction of [Su10].) One of the deep approximations Bogoliubov made was the $c$-number substitution of the operators creating and annihilating a particle of zero momentum; its justification was the subject of later papers [G5, Su8, Su9, Li6]. Another contribution of Bogoliubov, his inequality and $1/q^2$-theorem became the main tool to prove the absence of BEC -- and the breakdown of a continuous symmetry in general -- at positive temperatures in one- and two-dimensional quantum-mechanical models [Bog2].

Analytical methods permit to get closer to the problem of liquid helium. A popular one of the fifties was the use of pseudo-potentials [Hu3, Hu4]. However, the {\em par excellence} analytical method is functional integration -- the one we employed here and in the preceding papers [Su11, Su12]. Feynman devised it to solve the time-dependent Schr\"odinger equation and to compute $\left\langle x\left|e^{-it H/\hbar}\right| x\right\rangle$ [Fe1]. The mathematical justification for imaginary time $it=\hbar\beta$, where $\beta$ is real positive, as a functional integral with the Wiener measure came from Kac [Kac1,2]. The acknowledgement in [Fe2] reveals that it must have been Kac who convinced Feynman to apply what we call today the Feynman-Kac formula to the $\lambda$-transition of liquid helium. The result was three seminal papers [Fe2-4], the first of which was devoted to a first-principle study of the $\lambda$-transition. Although Feynman was not interested in a rigorous proof, he seized a point which gained importance with time: the appearance of long permutation cycles during the transition.

In 1954 London published his macroscopic theory of superfluidity [Lon3] whose emblematic element is a macroscopic wave function associated with the condensate. This and Landau's theory [Lan1,2] constitute the basis for the phenomenological description of superfluidity. Another cardinal contribution from the fifties was due to Oliver Penrose and Onsager [Pen]. These authors found the connection between the expected number of particles in the condensate and the largest eigenvalue of the one-particle reduced density matrix $\sigma^{N,L}_1$. For an alternative characterization of BEC they used the non-decay of the off-diagonal element of the integral kernel of $\sigma_1=\lim_{N,L\to\infty}\sigma^{N,L}_1$, which later was extended into the notion of off-diagonal long-range order [Ya]. They gave the first and surprisingly precise estimate of the condensate fraction in the ground state of liquid helium, $\sim 8\%$, only slightly modified later by a sophisticated numerical calculation [Ce] and deduction from experiment [Sn], [So]. They extended their study to positive temperatures and found a connection between BEC and the fraction of particles in "large" permutation cycles.

In 1960 the start of the Journal of Mathematical Physics marked the adulthood of a new discipline. While the theory of BEC seemed more or less settled for most physicists, their more math-minded colleagues saw there a field to explore. The mathematical results of this decade are mostly negative, they prove the absence of BEC in different situations: at low fugacity, in one and two dimensions at positive temperatures, and in one dimension in the ground state. In three thorough and difficult papers Ginibre adapted the method of combining Banach space technics with the Kirkwood-Salzburg equation [Bog3, R2, R4, R5, Bog4] to quantum statistics, and proved the existence, analyticity and exponential clustering of the reduced density matrices at low fugacity in the thermodynamic limit [G2-4], see also [G1]. This remains until today the most elaborate application of the path integral method in statistical physics. For the quantal version of Tonks' hard rod model [Ton] Girardeau established the ground state, easily deducible from that of the free Fermi gas [Gi]. However, even the explicit knowledge of the ground state did not permit to decide quickly about BEC; finally, it was shown not to exist [Schu, Len]. The analogous model with soft-delta interaction is more difficult, it was solved with Bethe Ansatz by Lieb and Liniger for the ground state [Li1] and by Lieb for the excited states [Li2]. There is probably no BEC in this model either, approximate methods predict an algebraic decay of the off-diagonal correlation [Hal, Cr, Ko]. The general one-dimensional case was studied much later, with the conclusion that neither diagonal nor off-diagonal long-range order can exist in the ground state provided that the compressibility is finite [Pi2]. The question of the breakdown of a continuous symmetry at positive temperatures in one and two dimensions was discussed in generality by Wagner [W]. BEC can be considered as an instance of such a breakdown: that of gauge invariance [Fa2, Su8, Su9, Li6]. Its absence was shown by Hohenberg [Ho], the completion of the proof with the extension of Bogoliubov's inequality to unbounded operators was done later [Bou].

The most important result of the seventies was the long-awaited first proof of BEC in a system of interacting bosons in three dimensions. This was done on the cubic lattice for hard-core bosons at half-filling by Dyson, Lieb and Simon [Dy]. The extension to the ground state on the square lattice took another decade [Ken, Kub].
In the seventies two schools, one in Leuven directed by Andr\'e Verbeure and another in Dublin under John Lewis' leadership started a research program in quantum statistical physics, with special emphasis on different aspects of BEC. The main tool of the Leuven School was operator algebraic. The Dublin School dominantly worked on BEC in the free Bose gas with different domain shapes, boundary conditions, imperfection, for bosons with spins, etc. Later this group pioneered the application of the Large Deviation Principle to problems of the Bose gas [Lew1, Ber2, Ber3, Ber4]. The intertwined activity of these two schools is reviewed in Verbeure's book [Ver].

In the mathematical literature of the interacting Bose gas most often either the pair potential or its Fourier transform is chosen to be nonnegative, see e.g. [Li7, Car, MP1, MP2, Su10]  -- in our theorem we needed both. Apart from the argument we gave in the Introduction, the estimates are easier, the bounds are sharper with a sign-keeping potential. Another reason is specific to dilute systems: the effect of a nonnegative spherical pair potential in the dilute limit reduces to $s$-wave scattering, so the interaction can be characterized by a single parameter, the $s$-wave scattering length $a$. This made it possible to obtain rigorously the ground state energy [Li3, Li4] and the free energy [Se1, Yi] in the dilute gas limit. Yet another question in which the positivity of the interaction played a role is the sign of the shift $\Delta T_c=T_c-T^0_c$ of the critical temperature in the dilute limit. Here $T_c$ and $T^0_c$ are the critical temperatures of the interacting and the ideal gas, respectively. There is a well-known argument saying that $r$-space repulsion implies $k$-space attraction [Hu1-Hu4]; so BEC should be easier for repulsive bosons, $T_c>T^0_c$. In a long debate a consensus has formed that the shift was positive, and $\Delta T_c/T^0_c\approx 1.3\sqrt{a\rho^{1/3}}$ in three dimensions. Although the question is pertinent, it cannot be decided without proving $T_c>0$, which was done only for the free Bose gas in a nonnegative external field, where $\Delta T_c>0$ follows from the min-max principle [Kac3]. A rigorous upper bound on $\Delta T_c$ and a review of the related literature can be found in [Se2]. The only case offering the possibility of a comparison with experiment is the $\lambda$-transition of liquid helium. Here the shift is negative [Lon2], but the system is dense, and the pair potential acting in it has an attractive tail.

Explicit summation over the symmetric group is part of the first-quantized treatment of quantum many-body systems, in particular when the Feynman-Kac formula is applied. The more elegant algebraic method has its limits, and from the early nineties there has been a substantial reappearance of the path integral technic in quantum statistical mechanics.
A paper by Aizenman and Lieb [Ai1] used it to prove the partial survival of Nagaoka ferromagnetism in the Hubbard model at positive temperatures. T\'oth [Tth] proved BEC of hard-core bosons on the complete graph. Aizenman and Nachtergaele [Ai2] studied ordering in the ground state of quantum spin chains. Ceperly [Ce] applied the path-integral Monte Carlo method to a thorough numerical analysis of the superfluidity of liquid helium. The present author picked up the thread left by Feynman and discussed BEC of particles in continuous space in connection with the probability distribution of permutation cycles [Su1, Su2].
The revival of interest in the relation between BEC and infinite cycles gave rise to many other papers, e.g. [Bun], [Scha], [Uel1], [Uel2], [Ben], [Do3], [Ad1]. There appeared also a new field of research on Hamiltonian models of random permutations, apparently more amenable to study by functional integration and large deviations analysis; see e.g. [Bet1], [Bet2], [Bet3], [El], [Ad2].

The path integral method can be adapted to systems of particles on a lattice [Ai1, Ai2], so our theorem applies to certain Bose-Hubbard models. Unfortunately, we were unable to treat hard-core interactions, therefore the most interesting of them, those of half spins are unavailable for us: the beautiful results obtained by reflection positivity for classical [Fr] and quantum [Dy] Heisenberg models still wait for an extension.

%\newpage
\vspace{20pt}
\noindent{\Large\bf References}
\begin{enumerate}
\item[{[Ad1]}] Adams S., Collevecchio A., and K\"onig W.: {\em A variational formula for the free energy of an interacting many-particle system.} Ann. Prob. {\bf 39}, 683-728 (2011).
\item[{[Ad2]}] Adams S. and Dickson M.: {\em An explicit large deviations analysis of the spatial cycle Huang-Yang-Luttinger model.} Ann. Henri Poincar\'e {\bf 22}, 1535-1560 (2021).
\item[{[Ai1]}] Aizenman M. and Lieb E. H.: {\em Magnetic properties of some itinerant-electron systems at $T>0$.} Phys. Rev. Lett. {\bf 65}, 1470-1473 (1990).
\item[{[Ai2]}] Aizenman M. and Nachtergaele B.: {\em Geometric aspects of quantum spin states.} Commun. Math. Phys. {\bf 164}, 17-63 (1994).
%\item[{[Ald]}] Alder B. J.and Wainwright T. E.: {\em Phase transition in elastic discs.} Phys. Rev. {\bf 127}, 359-361 (1962).
\item[{[All]}] Allen J. F. and Misener D.: {\em Flow of liquid helium II.} Nature {\bf 141},  75 (1938), and {\bf 142}, 643-644 (1938).
%\item[{[Az]}] Aziz R. A., Nain V. P. S., Carley J. S., Taylor W. L. and McConville G. T.: {\em An accurate intermolecular potential for helium.} J. Chem Phys. {\bf 70}, 4330-4342 (1979).
\item[{[Ben]}] Benfatto G., Cassandro M., Merola I. and Presutti E.: {\em Limit theorems for statistics of combinatorial partitions with applications to mean field Bose gas.} J. Math. Phys. {\bf 46}, 033303 (2005).
\item[{[Ber1]}] van den Berg M., Lewis J. T and de Smedt Ph: {\em Condensation in the imperfect boson gas.} J. Stat. Phys. {\bf 37}, 697-707 (1984).
\item[{[Ber2]}] van den Berg M., Lewis J. T. and Pul\'e J. V.: {\em The Large Deviation Principle and some models of an interacting Boson gas.} Commun. Math. Phys. {\bf 118}, 61-85 (1988).
\item[{[Ber3]}] van den Berg M., Dorlas T. C., Lewis J. T. and Pul\'e J. V.: {\em A perturbed mean field model of an interacting boson gas and the Large Deviation Principle.} Commun. Math. Phys. {\bf 127}, 41-69 (1990).
 \item[{[Ber4]}] van den Berg M., Dorlas T. C., Lewis J. T. and Pul\'e J. V.: {\em The pressure in the Huang-Yang-Luttinger model of an interacting Boson gas.} Commun. Math. Phys. {\bf 128}, 231-245 (1990).
\item[{[Bet1]}] Betz V. and Ueltschi D.: {\em Spatial random permutations and infinite cycles.} Commun. Math. Phys. {\bf 285}, 469-501 (2009).
\item[{[Bet2]}] Betz V. and Ueltschi D.: {\em Spatial random permutations and Poisson-Dirichlet law of cycle lengths.} Electr. J. Probab. {\bf 16}, 1173-1192 (2011).
\item[{[Bet3]}] Betz V., Ueltschi D. and Velenik Y.: {\em Random permutations with cycle weights.} Ann. Appl. Probab. {\bf 21}, 312-331 (2011).
%\item[{[Boe]}] de Boer J. and Michels A.: {\em Contribution to the quantum-mechanical theory of the equation of state and the law of corresponding states. Determination of the law of force of helium.} Physica {\bf 5}, 945-957 (1938).
\item[{[Bog1]}] Bogoliubov N. N.: {\em On the theory of superfluidity.} J. Phys. USSR {\bf 11}, 23-32 (1947).
\item[{[Bog2]}] Bogoliubov N. N.: {\em Quasi-averages in problems of statistical mechanics.} Dubna Report No. D-781, (1961), Ch. II (in Russian); Phys. Abhandl. Sowijetunion, 1962, {\bf 6}, 1-110; ibid., 1962, {\bf 6}, 113-229 (in German); {\em Selected Works, vol.II: Quantum Statistical Mechanics.} Gordon and Breach (N.Y. 1991); Collection of Scientific Papers in 12 vols.: Statistical Mechanics, vol.6, Part II. Nauka (Moscow 2006).
\item[{[Bog3]}] Bogolyubov N. N. and Khatset B. I.: {\em On some mathematical problems of the theory of statistical equilibrium.} Dokl. Akad. Nauk SSSR {\bf 66}, 321 (1949).
\item[{[Bog4]}] Bogolyubov N. N., Petrina D. Ya. and Khatset B. I.: {\em Mathematical description of the equilibrium state of classical systems on the basis of the canonical ensemble formalism.} Teor. Mat. Fiz. {\bf 1}, 251-274 (1969) and Ukr. J. Phys. {\bf 53}, 168-184 (2008).
%\item[{[Bol]}] Bollob\'as B.: {\em A probabilistic proof of an asymptotic formula for the number of labelled regular graphs.} Europ. J. Combinatorics {\bf 1}, 311-316 (1980).
\item[{[Bos]}] Bose S. N.: {\em Plancks Gesetz und Lichtquantenhypothese.} Z. Phys. \textbf{26}, 178-181 (1924).
\item[{[Bou]}] Bouziane M. and Martin Ph. A.: {\em Bogoliubov inequality for unbounded operators and the Bose gas.} J. Math. Phys. {\bf 17}, 1848-1851 (1976).
%\item[{[Bow]}] Bowen L., Lyons R., Radin Ch. and Winkler P.: {\em Fluid-solid transition in a hard-core system.} Phys. Rev. Lett. {\bf 96}, 025701 (2006).
%\item[{[Buc]}] Buckingham R. A.: {\em The classical equation of state of gaseous helium, neon and argon.} Proc. R. Soc. {\bf A168}, 264-283 (1938).
\item[{[Buf]}] Buffet E. and Pul\'e J. V.: {\em Fluctuation properties of the imperfect Bose gas.} J. Math. Phys. {\bf 24}, 1608-1616 (1983).
\item[{[Bun]}] Bund S. and Schakel M. J.: {\em String picture of Bose-Einstein condensation.} Mod. Phys. Lett. B {\bf 13}, 349 (1999).
\item[{[Car]}] Carlen E. A., Holzmann M., Jauslin I., Lieb E. H.: {\em A fresh look at a simplified approach to the Bose gas.} Physical Review A {\bf 103}, 053309 (2021).
\item[{[Ch]}] Choquet G.:  \emph{Diam\`etre transfini et comparaison de diverses capacit\'es.} S\'eminaire Brelot-Choquet-Deny. Th\'eorie du potentiel \textbf{3}, No. 4, 1-7 (1958-1959).
\item[{[Ce]}] Ceperley D. M.: {\em Path integrals in the theory of condensed helium.} Rev. Mod. Phys. {\bf 67}, 279-355 (1995).
\item[{[Cr]}] Creamer D. B., Thacker H. B. and Wilkinson D.: {\em A study of correlation functions for the delta-function Bose gas.} Physica {\bf 20D}, 155-186 (1986).
\item[{[Dav]}] Davies E. B.: {\em The thermodynamic limit for an imperfect boson gas.} Commun. Math. Phys. {\bf 28}, 69-86 (1972).
\item[{[Do1]}] Dorlas T. C., Lewis J. T. and Pul\'e J. V.: {\em Condensation in some perturbed meanfield models of a Bose gas.} Helv. Phys. Acta {\bf 64}, 1200-1224 (1991).
\item[{[Do2]}] Dorlas T. C., Lewis J. T. and Pul\'e J. V.: {\em The full diagonal model of a Bose gas.} Commun. Math. Phys. {\bf 156}, 37-65 (1993).
\item[{[Do3]}] Dorlas T. C., Martin Ph. A. and Pul\'e J. V.: {\em Long cycles in a perturbed mean field model of a boson gas.} J. Stat. Phys. {\bf 121}, 433-461 (2005).
\item[{[Dy]}] Dyson F. J., Lieb E. H. and Simon B.: {\em Phase transitions in quantum spin systems with isotropic and nonisotropic interactions.} J. Stat. Phys. {\bf 18}, 335-383 (1978).
\item[{[E1]}] Einstein A.: {\em Quantentheorie des einatomigen idealen Gases.} Sitz.ber. Preuss. Akad. Wiss. {\bf 1924}, 261-267.
\item[{[E2]}] Einstein A.: {\em Quantentheorie des einatomigen idealen Gases. II.} Sitz.ber. Preuss. Akad. Wiss. {\bf 1925}, 3-14.
\item[{[E3]}] Einstein A.: {\em Quantentheorie des idealen Gases.} Sitz.ber. Preuss. Akad. Wiss. {\bf 1925}, 18-25.
\item[{[E4]}] Einstein A.: {\em Answer to Schr\"odinger on 28 February 1925.} The collected papers of Albert Einstein, Vol. 14, Document 446, p. 438.
\item[{[El]}] Elboim D. and Peled R.: {\em Limit distributions for Euclidean random permutations.} Commun. Math. Phys. {\bf 369}, 457-522 (2019).
\item[{[Fa1]}] Fannes M. and Verbeure A.: {\em The condensed phase of the imperfect Bose gas.} J. Math. Phys. {\bf 21}, 1809-1818 (1980).
\item[{[Fa2]}] Fannes M., Pul\'e J. V. and Verbeure A.: {\em On Bose condensation.} Helv. Phys. Acta {\bf 55}, 391-399 (1982).
\item[{[Fe1]}] Feynman R. P.: {\em Space-time approach to non-relativistic quantum mechanics.} Rev. Mod. Phys. {\bf 20}, 367-387 (1948).
\item[{[Fe2]}] Feynman R. P.: {\em Atomic theory of the $\lambda$ transition in helium.} Phys. Rev. {\bf 91}, 1291-1301 (1953).
\item[{[Fe3]}] Feynman R. P.: {\em Atomic theory of liquid helium near absolute zero.} Phys. Rev. {\bf 91}, 1301-1308 (1953).
\item[{[Fe4]}] Feynman R. P.: {\em Atomic theory of the two-fluid model of liquid helium.} Phys. Rev. {\bf 94}, 262-277 (1954).
%\item[{[Fi]}] Fisher M. E. and Essam J. W.: {\em Some cluster size and percolation problems.} J. Math. Phys. {\bf 2}, 609-619 (1961).
\item[{[Fr]}] Fr\"ohlich J., Simon B. and Spencer T.: {\em Infrared bounds, phase transitions and continuous symmetry breaking.} Commun. Math. Phys. {\bf 50}, 79-85 (1976).
\item[{[G1]}] Ginibre J.: {\em Some applications of functional integration in Statistical Mechanics.} In: {\em Statistical Mechanics and Quantum Field Theory}, eds. C. De Witt and R. Stora, Gordon and Breach (New York 1971).
\item[{[G2]}] Ginibre J.: {\em Reduced density matrices of quantum gases. I. Limit of infinite volume.} J. Math. Phys. {\bf 6}, 238-251 (1965).
\item[{[G3]}] Ginibre J.: {\em Reduced density matrices of quantum gases. II. Cluster property.} J. Math. Phys. {\bf 6}, 252-262 (1965).
\item[{[G4]}] Ginibre J.: {\em Reduced density matrices of quantum gases. III. Hard-core potentials.} J. Math. Phys. {\bf 6}, 1432-1446 (1965).
\item[{[G5]}] Ginibre J.: {\em On the asymptotic exactness of the Bogolyubov approximation for many boson systems.} Commun. Math. Phys. {\bf 8}, 26-51 (1968).
\item[{[Gi]}] Girardeau M.: {\em Relationship between systems of impenetrable bosons and fermions in one dimension.} J. Math. Phys. {\bf 1}, 516-523 (1960).
%\item[{[Gl]}] Glyde H. R.: {\em Solid Helium. In: Rare gas solids, vol. I}, eds. M. L. Klein and J. A. Venables, Academic Press (London-New York-San Francisco, 1976), fig. 1.
\item[{[Gr]}] Griffin A.: {\em Excitations in a Bose-condensed liquid} (Cambridge University Press, 1993).
\item[{[Hal]}] Haldane F. D. M.: {\em Effective harmonic-fluid approach to low-energy properties of one-dimensional quantum fluids.} Phys. Rev. Lett. {\bf 47}, 1840-1843 (1981).
%\item[{[Han]}] Hansen J.-P. and Verlet L.: {\em Phase transitions of the Lennard-Jones system.} Phys. Rev. {\bf 184}, 151-161 (1969).
\item[{[Ho]}] Hohenberg P. C.: {\em Existence of long-range order in one and two dimensions.} Phys. Rev. {\bf 158}, 383-386 (1967).
\item[{[Hu1]}] Huang K.: {\em Imperfect Bose gas.} In: {\em Studies in Statistical Mechanics, Vol II.} eds. J. de Boer and G. E. Uhlenbeck, North-Holland (Amsterdam 1964), pp. 1-106.
\item[{[Hu2]}] Huang K.: {\em Statistical Mechanics.} 2nd ed. Wiley (New York 1987), p. 303, Prob. 12.7.
\item[{[Hu3]}] Huang K. and Yang C. N.: {\em Quantum-mechanical many-body problem with hard-sphere interaction.} Phys. Rev. {\bf 105}, 767-775 (1957).
\item[{[Hu4]}] Huang K., Yang C. N. and Luttinger J. M.: {\em Imperfect Bose with hard-sphere interactions.} Phys. Rev. {\bf 105}, 776-784 (1957).
%\item[{[J]}] Johnston D. C.: {\em Thermodynamic properties of the van der Waals fluid.} arXiv:1402.1205 (2014) Fig. 7.
\item[{[Kac1]}] Kac M.: {\em On distributions of certain Wiener functionals.} Trans. Amer. Math. Soc. {\bf 65}, 1-13 (1949).
\item[{[Kac2]}] Kac M.: {\em On some connections between probability theory and differential and integral equations.} In: Proceedings of the Second Berkeley Symposium on Probability and Statistics, J. Neyman ed., Berkeley, University of California Press (1951).
\item[{[Kac3]}] Kac M. and Luttinger J. M.: {\em Bose-Einstein condensation in the presence of imputities.} J. Math. Phys. {\bf 14}, 1626-1628 (1973).
\item[{[Kad]}] Kadanoff L. P.: {\em Slippery wave functions.} J. Stat. Phys. {\bf 152}, 805-823 (2013).
\item[{[Kag]}] Kagan M. Yu.. {\em Modern trends in superconductivity and superfluidity} (Springer-Verlag, 2013).
\item[{[Kah]}] Kahn B. and Uhlenbeck G. E.: {\em On the theory of condensation.} Physica {\bf 5}, 399-416 (1938).
\item[{[Kap]}] Kapitza P.: {\em Viscosity of liquid helium below the $\lambda$-point.} Nature {\bf 141}, 74 (1938).
\item[{[Ken]}] Kennedy T., Lieb E. H. and Shastry B. S.: {\em The XY model has long-range order for all spins and all dimensions greater than one.} Phys. Rev. Lett. {\bf 61}, 2582-2584 (1988).
%\item[{[Kim]}] Kim E. and Chan M. H. W.: {\em Observation of superflow in solid helium.} Science {\bf 305}, 1941-1944 (2004).
\item[{[Ko]}] Korepin V. E., Bogoliubov N. M. and Izergin A. G.: {\em Quantum inverse scattering method and correlation functions.} Cambridge University Press (1993) Ch. XVIII.2.
\item[{[Kub]}] Kubo K. and Kishi T.: {\em Existence of long-range order in the XXZ model.} Phys. Rev. Lett. {\bf 61}, 2585-2587 (1988).
%\item[{[Kun]}] Kunz H. and Souillard B.: {\em Essential singularity in percolation problems and asymptotic behavior of cluster size distribution.} J. Stat. Phys. {\bf 19}, 77-106 (1978).
\item[{[Lan1]}] Landau L. D.: {\em The theory of superfluidity of helium II.} J. Phys. USSR {\bf 5}, 71-90 (1941).
\item[{[Lan2]}] Landau, L.D.: {\em On the theory of supefluidity of helium II.} J. Phys. USSR {\bf 11}, 91-92 (1947).
%\item[{[Leb]}] Lebowitz J. L., Mazel A. and Presutti E.: {\em Liquid-vapor phase transitions for systems with finite-range interactions.} J. Stat. Phys. {\bf 94}, 955-1025 (1999).
\item[{[Leg]}] Leggett A. J.: {\em Quantum liquids} (Oxford University Press, 2006).
\item[{[Len]}] Lenard A.: {\em Momentum distribution in the ground state of the one-dimensional system of impenetrable bosons.} J. Math. Phys. {\bf 5}, 930-943 (1964).
%\item[{[Lenn]}] Lennard-Jones J. E.: {\em On the determination of molecular fields.} Proc. Roy. Soc. Lond. A{\bf 106}, 463-477 (1924).
\item[{[Lew1]}] Lewis J. T., Zagrebnov V. A. and Pul\'e J. V.: {\em The large deviation principle for the Kac distribution.} Helv. Phys. Acta {\bf 61} 1063-1078 (1988).
%\item[{[Lew2]}] Lewis J. T., Pfister C.-E. and Sullivan W. G.: {\em Large deviations and the thermodynamic formalism: A new proof of the equivalence of ensembles.} In \emph{"On three levels"}, eds. M. Fannes, Ch. Maes, A. Verbeure (Springer 1994) pp. 183-192.
\item[{[Li1]}] Lieb E. H. and Liniger W.: {\em Exact analysis of an interacting Bose Gas. I. The general solution and the ground state.} Phys. Rev. {\bf 130}, 1605-1616 (1963).
\item[{[Li2]}] Lieb E. H.: {\em Exact analysis of an interacting Bose gas. II. The excitation spectrum.} Phys. Rev. {\bf 130}, 1616-1624 (1963).
\item[{[Li3]}] Lieb E. H. and Yngvason J.: {\em Ground state energy of the low density Bose gas.} Phys. Rev. Lett. {\bf 80}, 2504-2507 (1998).
\item[{[Li4]}] Lieb E. H. and Yngvason J.: {\em The ground state energy of a dilute two-dimensional Bose gas.} J. Stat. Phys. {\bf 103}, 509-526 (2001).
\item[{[Li5]}] Lieb E. H., Seiringer R., Solovey J. P. and Yngvason J.: {\em The mathematics of the Bose gas and its condensation.}  Birkh\"auser Verlag (Basel-Boston-Berlin 2005), and arXiv:cond-mat/0610117.
\item[{[Li6]}] Lieb E. H., Seiringer R. and Yngvason J.: {\em Justification of {\rm c}-number substitution in bosonic Hamiltonians.} Phys. Rev. Lett. {\bf 94}, 080401 (2005).
\item[{[Li7]}]  Lieb E. H.: {\em  Simplified approach to the ground-state energy of an imperfect Bose gas.} Phys. Rev. {\bf 130}, 2518-2528 (1963).
%\item[{[Lin]}] Van der Linden J.: {\em On the asymptotic problem of statistical thermodynamics for a real system. III. Questions concerning chemical potential and pressure.} Physica {\bf 38}, 173-188 (1968).
\item[{[Lon1]}] London F.: {\em The $\lambda$-phenomenon of liquid helium and the Bose-Einstein degeneracy.} Nature {\bf 141}, 643-644 (1938).
\item[{[Lon2]}] London F.: {\em On the Bose-Einstein condensation.} Phys. Rev. {\bf 54}, 947-954 (1938).
%\item[{[Lon3]}] London F.: {\em Zur Theorie und Systematik der Molekularkr\"afte.} Z. Phys. {\bf 63}, 245-279 (1930).
\item[{[Lon3]}] London F.: {\em Superfluids, Vol. 2: Macroscopic theory of superfluid helium.} (Wiley, New York, 1954).
%\item[{[M]}] Mie G.: {\em Zur kinetischen Theorie der einatomigen K\"orper.} Annalen der Physik \textbf{316}, 657-697 (1903).
\item[{[MP1]}] Martin Ph. A. and Piasecki J.: {\em Self-consistent equation for an interacting Bose gas.} Phys. Rev. E {\bf 68}, 016113 (2003).
\item[{[MP2]}] Martin Ph. A. and Piasecki J.: {\em Bose gas beyond mean field .} Phys. Rev. E {\bf 71}, 016109 (2005).
%\item[{[Ni]}] Niebel K. F. and Venables J. A.: {\em The crystal structure problem. In: Rare gas solids, vol. I}, eds. M. L. Klein and J. A. Venables, Academic Press (London-New York-San Francisco, 1976).
\item[{[No]}] Nozi\`eres P. and Pines D.: {\em Theory of quantum liquids II: Superfluid Bose liquids.} Addison-Wesley (Redwood City, 1990).
\item[{[Pen]}] Penrose O. and Onsager L.: {\em Bose-Einstein condensation and liquid He.} Phys. Rev. {\bf 104}, 576-584 (1956).
\item[{[Pet]}] Pethick C. J. and Smith H.: {\em Bose-Einstein condensation in dilute gases.} Cambridge University Press (2002).
\item[{[Pi1]}] Pitaevskii L. and Stringari S.: {\em Bose-Einstein condensation.} Clarendon Press (Oxford 2003).
\item[{[Pi2]}] Pitaevskii L. and Stringari S.: {\em Uncertainty principle, quantum fluctuations, and broken symmetries.} J. Low Temp. Phys. {\bf 85}, 377-388 (1991).
\item[{[R1]}] Ruelle D.: {\em Classical statistical mechanics of a system of particles.} Helv. Phys. Acta \textbf{36}, 183-197 (1963).
\item[{[R2]}] Ruelle D.: \emph{Statistical Mechanics.} W. A. Benjamin (New York-Amsterdam 1969).
\item[{[R3]}] Ruelle D.: {\em Superstable interactions in classical statistical mechanics.} Commun. Math. Phys. {\bf 18}, 127-159 (1970).
\item[{[R4]}] Ruelle D.: {\em Correlation functions of classical gases.} Ann. Phys. {\bf 25}, 109-120 (1963).
\item[{[R5]}] Ruelle D.: {\em Cluster property of the correlation functions of classical gases.} Rev. Mod. Phys. {\bf 36}, 580-584 (1964).
\item[{[Scha]}] Schakel A. M. J.: {\em Percolation, Bose-Einstein condensation, and string proliferation.} Phys. Rev. E {\bf 63}, 026115 (2001).
\item[{[Schr]}] Schr\"odinger E.: {\em Letter to Einstein on 5 February 1925.} The collected papers of Albert Einstein, Vol. 14, Document 433, p. 429.
\item[{[Schu]}] Schultz T. D.: {\em Note on the one-dimensional gas of impenetrable point-particle bosons.} J. Math. Phys. {\bf 4}, 666-671 (1963).
\item[{[Se1]}] Seiringer R.: {\em Free energy of a dilute Bose gas: Lower bound.} Commun. Math. Phys. {\bf 279}, 595-636 (2008).
\item[{[Se2]}] Seiringer R. and Ueltschi D.: {\em Rigorous upper bound on the critical temperature of dilute Bose gases.} Phys. Rev. B {\bf 80}, 014502 (2009).
\item[{[Sew]}] Sewell G. L.: {\em Quantum mechanics and its emergent macrophysics.} Princeton University Press (2002).
%\item[{[Sl]}] Slater J. C. and Kirkwood J. G.: {\em The van der Waals forces in gases.} Phys. Rev. {\bf 37}, 682-697 (1931).
\item[{[Sn]}] Snow W. M. and Sokol P. E.: {\em Density and temperature dependence of the momentum distribution in liquid Helium 4.} J. Low. Temp. Phys. {\bf 101}, 881-928 (1995).
\item[{[So]}] Sosnick T. R., Snow W. M. and Sokol P. E.: {\em Deep-inelastic neutron scattering from liquid He4.} Phys. Rev. B{\bf 41}, 11185-11202 (1990).
\item[{[Su1]}] S\"ut\H o A.: {\em Percolation transition in the Bose gas.} J. Phys. A: Math. Gen. {\bf 26}, 4689-4710 (1993).
\item[{[Su2]}] S\"ut\H o A.: {\em Percolation transition in the Bose gas: II.} J. Phys. A: Math. Gen. {\bf 35}, 6995-7002 (2002). See also arXiv:cond-mat/0204430v4 with addenda after Eqs.~(34) and (44).
%\item[{[Su3]}] S\"ut\H o A.: {\em Thermodynamic limit and proof of condensation for trapped bosons.} J. Stat. Phys. {\bf 112}, 375-396 (2003).
%\item[{[Su4]}] S\"ut\H o A.: {\em Correlation inequalities for noninteracting Bose gases.} J. Phys. A: Math. Gen. {\bf 37}, 615-621 (2004).
\item[{[Su5]}] S\"ut\H o A.: {\em Ground state at high density.} Commun. Math. Phys. {\bf 305}, 657-710 (2011).
%\item[{[Su6]}] S\"ut\H o A.: {\em The total momentum of quantum fluids.} J. Math. Phys. \textbf{56}, 081901 (2015), Section IV.
%\item[{[Su7]}] S\"ut\H o A.: {\em A possible mechanism of concurring diagonal and off-diagonal long-range order for soft interactions.} J. Math. Phys. \textbf{50}, 032107 (2009).
\item[{[Su8]}] S\"ut\H o A.: {\em Equivalence of Bose-Einstein condensation and symmetry breaking.} Phys. Rev. Lett. {\bf 94}, 080402 (2005).
\item[{[Su9]}] S\"ut\H o A.: {\em Bose-Einstein condensation and symmetry breaking.} Phys. Rev. A {\bf 71}, 023602 (2005).
\item[{[Su10]}] S\"ut\H o A. and Sz\'epfalusy P.: {\em Variational wave functions for homogenous Bose systems.} Phys. Rev. {\bf A77}, 023606 (2008).
\item[{[Su11]}] S\"ut\H o A.: {\em Fourier formula for quantum partition functions.} arXiv:2106.10032 [math-ph] (2021).
\item[{[Su12]}] S\"ut\H o A.: {\em Simultaneous occurrence of off-diagonal long-range order and infinite permutation cycles in systems of interacting atoms.} arXiv:2108.02659 [math-ph] (2021).
\item[{[Su13]}] S\"ut\H o A.: {\em Crystalline ground states for classical particles.} Phys. Rev. Lett. {\bf 95}, 265501 (2005).
\item[{[T1]}] Tisza L.: {\em Transport phenomena in helium II.} Nature {\bf 141}, 913 (1938).
\item[{[T2]}] Tisza L.: {\em La viscosit\'e de h\'elium liquide et la statistique de Bose-Einstein.} C. R. Paris {\bf 207}, 1035-1186 (1938).
%\item[{[Toe]}] Toennies J. P. and Vilesov A. F.: {\em Superfluid helium droplets: A uniquely cold nanomatrix for molecules and molecular complexes.} Angew. Chem. Int. Ed. {\bf 43}, 2622-2648 (2004) Table 2.
\item[{[Ton]}] Tonks L.: {\em The complete equation of state of one, two and three-dimensional gases of hard elastic spheres.} Phys. Rev. {\bf 50}, 955-963 (1936).
%\item[{[Tot]}] T\'oth B.: {\em Improved lower bound on the thermodynamic pressure of the spin 1/2 Heisenberg ferromagnet.} Lett. Math. Phys. {\bf 28}, 75-84 (1993).
\item[{[Tth]}] T\'oth B.: {\em Phase transition in an interacting Bose system. An application of the theory of Ventsel’ and Freidlin.} J. Stat. Phys {\bf 61}, 749–764 (1990).
\item[{[Ued]}] Ueda M.: {\em Fundamentals and new frontiers of Bose-Einstein condensation.} World Scientific (2010).
\item[{[Uel1]}] Ueltschi D.: {\em Relation between Feynman cycles and off-diagonal long-range order.} Phys. Rev. Lett. {\bf 97}, 170601 (2006).
\item[{[Uel2]}] Ueltschi D.: {\em Feynman cycles in the Bose gas.} J. Math. Phys. {\bf 47}, 123303 (2006).
\item[{[Uh]}] Uhlenbeck G. E.: {\em Dissertation Leiden} 1927, p. 69.
\item[{[Ver]}] Verbeure A.: {\em Many body boson systems: half a century later.} Springer (2011).
\item[{[Vers]}] Vershik A. M.: {\em Statistical mechanics of combinatorial partitions, and their limit shapes.} Funk. Anal. Appl. {\bf 30}, 90-105 (1996).
\item[{[W]}] Wagner H.: {\em Long-wavelength excitations and the Goldstone theorem in many-particle systems with "broken symmetries".} Z. Phys. {\bf 195}, 273-299 (1966).
\item[{[Ya]}] Yang C. N.: {\em Concept of off-diagonal long-range order and the quantum phases of liquid He and of superconductors.} Rev. Mod. Phys. {\bf 34}, 694-704 (1962).
\item[{[Yi]}] Yin J.: {\em Free energies of dilute Bose gases: Upper bound.} J. Stat. Phys. {\bf 141}, 683-726 (2010).
\item[{[Z]}] Zagrebnov V. A. and Bru J.-B.: {\em The Bogoliubov model of weakly imperfect Bose gas.} Phys. Rep. {\bf 350}, 291-434 (2001).
\end{enumerate}
\end{document}